%% file: Main.tex
\documentclass[journal]{IEEEtran}

\usepackage{amsmath,amsfonts}

\usepackage{mathrsfs}

\usepackage{algpseudocode}
\usepackage{algorithm}

\usepackage{textcomp}
\usepackage{stfloats}
\usepackage{url}
\usepackage{verbatim}
\usepackage{graphicx}
\usepackage{cite}

\usepackage[caption=false,font=footnotesize,labelformat=simple]{subfig}

\makeatletter

\renewcommand{\p@subfigure}{\thefigure} 
\makeatother

\usepackage{cuted}     
\usepackage{lipsum}    

\usepackage{booktabs}
\usepackage{tabularx, multirow, amssymb, textcomp, array}
\usepackage{colortbl} 

\usepackage{titlesec}

\begin{document}
{

\title{Agentic AI–Driven UAV Network Deployment$:$ An LLM–Enhanced Exact Potential Game Approach}

\author{Xin Tang, Xiaohuan Li, ~\IEEEmembership{Member,~IEEE}, Qian Chen, Binhan Liao, Yaqi Zhang, Jianxin Chen, Changyuan Zhao, Junchuan Fan, Junxi Tian 
        %
\IEEEcompsocitemizethanks{
\IEEEcompsocthanksitem X. Tang, B. Liao, Y. Zhang, J. Chen, J. Fan, and J. Tian are with the Guangxi University Key Laboratory of Intelligent Networking and Scenario System (School of Information and Communication, Guilin University of Electronic Technology (GUET)), Guilin, 541004, China (e-mails:tangx@mail.guet.edu.cn; guetlbh@126.com; zhangyaqi8@163.com; chenjianxin792@gmail.com; fanjunchuan@mails.guet.edu.cn; tech.tjx@hotmail.com).

\IEEEcompsocthanksitem X. Li is with the Guangxi University Key Laboratory of Intelligent Networking and Scenario System (School of Information and Communication, GUET), Guilin, 541004, China, and also with National Engineering Laboratory for Comprehensive Transportation Big Data Application Technology (Guangxi), Nanning, 530001, China (e-mail: lxhguet@guet.edu.cn).

\IEEEcompsocthanksitem Q. Chen is with the School of Architecture and Transportation Engineering, GUET, Guilin, 541004, China (e-mail: chenqian@mails.guet.edu.cn).

\IEEEcompsocthanksitem C. Zhao is with the College of Computing and Data Science, Nanyang Technological University, Singapore, and CNRS@CREATE, 1 Create Way, 08-01 Create Tower, Singapore 138602 (e-mail: zhao0441@e.ntu.edu.sg).

}

\thanks{This work was supported in part by the Guangxi Natural Science Foundation of China under Grant 2026GXNSFGB00640004 and 2025GXNSFAA069687, in part by the National Natural Science Foundation of China under Grant U22A2054. {\itshape (Corresponding author: Xiaohuan Li).}}}

\markboth{IEEE Transactions on Cognitive Communications and Networking,~Vol.~X, No.~X, X~2026}%
{Shell \MakeLowercase{\textit{et al.}}: Bare Advanced Demo of IEEEtran.cls for IEEE Computer Society Journals}

\IEEEtitleabstractindextext{%
\begin{abstract}
Unmanned aerial vehicular network (UAVN) is envisioned to provide flexible connectivity, wide-area coverage, and low-latency services in dynamic environments. From an agentic artificial intelligence (Agentic AI) perspective, UAVNs naturally operate as multi-agent systems, where UAVs act as intelligent agents that coordinate deployment and networking decisions to achieve global performance objectives. However, the strong coupling between discrete link decisions and continuous deployment parameters makes UAVN deployment optimization a mixed-integer nonconvex problem, resulting in challenges in scalability, efficiency, and solution consistency under dynamic network conditions. This paper proposes a dual spatial-scale UAVN deployment optimization framework based on exact potential games (EPGs), enhanced by Agentic AI. At the large spatial scale, a log-linear learning based EPG (L3-EPG) algorithm is developed to optimize inter-UAV link configurations, enabling sparse yet connected network topologies while reducing redundant links and interference. At the small spatial scale, an approximate gradient based EPG (AG-EPG) algorithm jointly optimizes UAV deployment, transmission power allocation, and ground user (GU) association to improve network throughput and latency. To further enhance adaptability across heterogeneous scenarios, a large language model (LLM) is incorporated as a knowledge-driven decision enhancer to automatically generate utility weights according to network characteristics, alleviating reliance on manual parameter tuning. Simulation results demonstrate that the proposed framework consistently outperforms baseline methods in terms of energy consumption, end-to-end latency, and system throughput.
\end{abstract}

}

\maketitle

\IEEEdisplaynontitleabstractindextext

\IEEEpeerreviewmaketitle

\ifCLASSOPTIONcompsoc
\IEEEraisesectionheading{
\section{Introduction}
\label{sec:introduction}}
\else
\section{Introduction}
\label{sec:introduction}
\fi
\IEEEPARstart{A}{gentic} artificial intelligence (Agentic AI) has recently emerged as a promising paradigm for building autonomous and goal-oriented intelligent systems. Unlike conventional data-driven AI models that focus on perception or prediction, Agentic AI emphasizes decision-making, action execution, and long-term objective optimization through interactions among multiple agents \cite{zhang2026toward}. Such systems typically consist of agents equipped with sensing, computation, and communication capabilities, enabling them to adapt their behaviors according to environmental dynamics and task requirements. The unmanned aerial vehicle network (UAVN) naturally serves as an important physical embodiment of Agentic AI in wireless communication systems. In UAVNs, each UAV operates as an agent that can perceive local network conditions, make topology and resource-related decisions, and cooperate with other UAVs to support network-level objectives. Owing to their high mobility, low deployment cost, and flexible configurability, UAVNs are increasingly regarded as a key component of emerging air–ground integrated communication networks \cite{li2023multi}. 

Despite these advantages, enabling effective agentic intelligence in UAVNs remains highly challenging. UAVs are subject to strict energy constraints, which limit network endurance and operational continuity \cite{ tang2025task}. Moreover, air–ground communication links are vulnerable to blockage and interference, making it difficult to maintain reliable connectivity \cite{zhang2024interactive}. In addition, UAVNs typically operate in highly dynamic environments and must jointly optimize multiple performance metrics, such as coverage, system throughput, energy consumption, and end-to-end latency. These factors collectively result in a highly complex and multi-objective optimization problem for UAVNs.


{\itshape \textbf{Challenge 1}: Bottlenecks in mixed integer non-convex modeling and centralized solving for UAVN deployment optimization.} 
UAVN deployment optimization is typically modeled as a Mixed Integer Nonlinear Programming (MINLP) problem \cite{zhan2024tradeoff}, making it difficult to obtain a globally optimal solution using precise methods \cite{sun2024joint}. Although convex relaxation methods \cite{wang2024energy}, centralized linear solvers \cite{gaydamaka_2024}, and nonlinear programming approaches \cite{zhang2024joint} can provide approximate solutions for small-scale static problems, they incur high computational and communication overhead. Moreover, their centralized nature introduces single-point failures, making them unsuitable for dynamic distributed UAVNs \cite{he2023location}.

{\itshape \textbf{Challenge 2}: Insufficient efficiency and deployability of heuristic and learning-based methods in UAVNs.} 
To alleviate the complexity of centralized optimization, researchers have proposed heuristic methods such as genetic algorithms\cite{gao2024joint}, particle swarm optimization\cite{wu2025novel}, and simulated annealing\cite{wan2024hybrid}. However, these methods usually rely on global information, have high computational complexity, are prone to getting trapped in local optima, and lack convergence guarantees\cite{wang2024bi}. Although UAVN deployment methods based on learning have certain environmental adaptability\cite{fu2023joint, chen2023deep, dai2023multi}, they rely on a large number of interaction samples and training computations, and the feasibility of actual network deployment is still limited \cite{gong2023bayesian}.

{\itshape \textbf{Challenge 3}: Limited global consistency and generalization ability in distributed optimization.} 
Game theory is widely used in UAVN optimization because it is suitable for describing distributed interactions between multiple agents \cite{du_2025}, and it has achieved certain results \cite{lv2023multi, liu2024uav, meng2023throughput}. Most game models are difficult to optimize for global utility and are prone to getting trapped in local optima \cite{li2024reinforcement}. By associating individual utility changes with the overall situation function, potential games can guarantee the convergence and consistency of the game \cite{zhang2024decentralized}. However, its adaptability and generalization ability in complex multi-scenario environments are limited, and it still relies on manual parameter tuning, making network optimization difficult.

Agentic AI in this paper refers to the high-level functional architecture that enables autonomous planning, goal-oriented task decomposition, and multi-agent coordination. It defines how the UAVN acts as a goal-seeking system to solve complex optimization problems. In contrast, the large language model (LLM) serves as a specific reasoning engine and knowledge source within Agentic AI. While Agentic AI provides the logic of agency and decision-making structures, the LLM facilitates semantic understanding of network intents and performs zero-shot reasoning for parameter generation based on the retrieved domain knowledge. In essence, Agentic AI is the framework of autonomous behavior, whereas the LLM is the cognitive tool that empowers the agents with expert-level reasoning capabilities.
This paper proposes an Agentic AI-driven UAVN deployment optimization approach based on large language model (LLM)-enhanced exact potential game (EPG). This method decomposes the complex MINLP into a dual spatial scale optimization for discrete links and continuous parameters. At large spatial scales, an EPG algorithm based on log-linear learning (L3-EPG) achieves a sparse configuration of network links through a probabilistic policy update mechanism, reducing redundant communication links and link interference while optimizing network connectivity \cite{li2024timestamp}. At small spatial scales, this paper proposes an approximate gradient based EPG (AG-EPG), which optimizes energy consumption, deployment, power, and user allocation of UAVNs in a distributed environment to improve system throughput and reduce latency. Both algorithms require only local information interaction among nodes to make decisions without relying on a central control node. Furthermore, to improve the adaptability and generalization ability of these algorithms in multiple scenarios and to reduce the complexity of manual parameter tuning, this paper introduces an LLM and a self-built knowledge base to automatically generate weights for utility functions. The main contributions of this paper are as follows:

\begin{itemize}
\item{
\textbf{Framework:} An Agentic AI-driven UAVN deployment model is constructed. This model comprehensively considers multiple optimization requirements, such as maximizing system throughput, minimizing UAV energy consumption, and minimizing end-to-end latency. An MINLP optimization problem is established involving discrete decision variables, such as the number of communication links between UAVs and the selection of ground user (GU) access associations, as well as continuous decision variables, such as the coordinates of the UAVs and the transmission power configuration. We employ Agentic AI to decompose high-level deployment objectives into structured sub-problems, where decision-making agents interact with the network environment and generate actions through game-based learning dynamics.
}

\item{
\textbf{Problem and decoupling:} To address the high-dimensional nonconvexity and NP-hardness of the MINLP, a dual spatial scale optimization strategy is proposed to solve the discrete link optimization and continuous parameter problems, respectively. An EPG theoretical model is constructed, designing a consistent mapping relationship between the local utility function and the global potential function. The existence, boundedness, and convergence of the potential function and the pure policy Nash equilibrium (NE) are rigorously proven. At a large spatial scale, an L3-EPG is proposed to effectively eliminate redundant links while ensuring network connectivity. At a small spatial scale, an AG-EPG algorithm is proposed to achieve joint optimization of UAV coordinate , transmission power allocation, and GU load balancing.
}

\item{
\textbf{Solution:} A dedicated domain knowledge base for UAVN deployment optimization is constructed, covering wireless communication theory, basic principles of game theory, and related optimization cases. Combining the Retrieval-augmented Generation (RAG), LLM is used to retrieve relevant knowledge and generate weight coefficients in the utility function based on specific feature parameters, such as node size, user distribution characteristics, and the degree of environmental occlusion in the network scenario. This effectively overcomes the limitations of traditional methods that heavily rely on manual experience for parameter tuning, significantly improving the algorithm's adaptability and generalization ability under different network configurations.
}

\item{
\textbf{Validation:} Extensive experiments are conducted to verify the performance of LLM-based knowledge retrieval, the convergence of the EPG, and deployment optimization, and to demonstrate the substantial advantages of the proposed approach over baseline algorithms in terms of energy consumption, end-to-end latency, and throughput.
}

\end{itemize}

The remainder of this paper is structured as follows. Section \ref{sec:2} reviews the related literature. Section \ref{sec:3} describes the system model. Section \ref{sec:4} formulates the optimization problem and further decomposes it into tractable subproblems. Section \ref{sec:5} develops an EPG based optimization framework for UAVN. Section \ref{sec:6} introduces an LLM enhanced utility function design and weight generation method. Section \ref{sec:7} presents the simulation results and performance analysis, followed by concluding remarks in Section \ref{sec:8}.

\section{Related Work}
\label{sec:2}
In this section, we review the related work on UAVN deployment optimization, Heuristic algorithm-based network optimization, and AI-based network optimization.

\subsection{UAVN deployment optimization}
Considerable research has been dedicated to the application of UAVs in emergency communication networks. Their high mobility, flexible deployment, and strong adaptability render them vital resources for supporting diverse operations, such as serving as aerial base stations to improve the connectivity of ground-based wireless systems. The authors in \cite{gaydamaka_2024} proposed dynamic topology organization and maintenance algorithms for autonomous UAV swarms, focusing on distributed methods to maintain network connectivity and robustness. The authors in \cite{sadi_2020} studied optimal measurement strategies for linear measurement systems and applied them to UAVN topology prediction. In \cite{tian2025optimal}, the authors addressed the optimal deployment problem for target localization by considering range-dependent noise and minimizing the cost function derived from the A-optimal criterion. In \cite{yoo_sk_2021}, the author explored biological robustness to design reliable UAVN, proposing the bio-LINK scheme to make the network resilient to UAV node failures. Line-of-Sight (LoS) coverage in UAV-based THz communication networks was analyzed in \cite{polese_2023}, focusing on 3D visualization of wireless networks and emphasizing deployment optimization in high-frequency bands to improve coverage and performance. The authors in \cite{du_2025} introduced formation-aware UAVN based on game-theoretic distributed topology control, aiming to optimize self-organization performance in multi-hop UAVN.

However, existing studies predominantly focus on UAV navigation, communication technologies, and civilian applications. The system modeling in this field often neglects the optimization of network topology specific to low-altitude environments.

\subsection{Heuristic algorithm-based network optimization}

Heuristic algorithms have attracted significant research attention for enhancing the performance of UAVN. These algorithms, known for their computational efficiency and ability to handle complex, non-linear constraints, offer a practical approach to solving critical problems such as dynamic trajectory planning, resource allocation, and network topology control in dynamic aerial environments. 

The authors in \cite{wang2024bi} utilized bi-objective ant colony optimization to jointly minimize energy consumption and completion time in UAV-assisted systems. In \cite{wan2024hybrid}, a hybrid heuristic algorithm was designed to optimize multi-UAV routing paths for time-dependent data collection tasks. In \cite{pan2023resource}, a multi-objective optimization approach was applied to handle resource scheduling in UAV-aided device-to-device communication networks. The authors in \cite{wu2025novel} proposed a hybrid enhanced particle swarm optimization technique to solve complex UAV path planning problems. In \cite{zhang2024joint}, an elite-driven differential evolution algorithm was developed to jointly optimize the deployment and flight planning of multi-UAVs. The authors in \cite{sun2024joint} addressed the joint optimization problem of task offloading and resource allocation in aerial-terrestrial networks. However, these heuristic-based methods often rely on difficult-to-obtain global information and entail high computational complexity. 

Furthermore, due to their stochastic nature, they are prone to getting trapped in local optima and lack rigorous convergence guarantees. While some studies have explored game-theoretic models to decentralize the decision-making process, many existing models lack strict assurances for global utility, potentially resulting in suboptimal overall system performance. Moreover, many applied game-theoretic models lack strict assurances for global utility, potentially resulting in suboptimal overall system performance.

\subsection{AI-based network optimization}

AI-assisted network optimization has garnered considerable interest as a transformative approach for UAVN.
In \cite{lv2023multi}, the author proposed a Multi-Agent Reinforcement Learning (MARL) approach to optimize relay selection and power allocation, enhancing the robustness and anti-jamming capability of UAV swarm communication networks. 
In\cite{fu2023joint}, the author proposed a game theory-based Deep Reinforcement Learning (DRL) to jointly optimize transmit power and the deployment of UAV base stations, significantly improving downlink throughput. 
In \cite{lv2025large}, the author proposed an LLM-driven DRL framework, where LLMs are used to automatically generate reward functions and improve the learning efficiency of resource scheduling in heterogeneous UAVN. 
In \cite{zhang2024generative} and \cite{wang2024optimizing}, the authors proposed an interactive generative AI agent framework for satellite networks by integrating LLMs with RAG and a mixture-of-experts transmission mechanism.

However, the practical deployment of such AI methods, particularly DRL, faces significant hurdles. These include their heavy reliance on data-intensive and computationally expensive training with safe environmental interaction, poor cross-scenario generalization, and a high dependency on expert tuning, which collectively increase optimization complexity and hinder real-world applications.

\section{System Model}
\label{sec:3}
To comprehensively characterize this dynamic network, this section details the system from five aspects: the network topology model, the UAV-to-UAV (U2U) and UAV-to-GU (U2G) communication model, the UAV energy consumption model, the network latency model, and the 3D mobility model of the UAVs.
\begin{figure}[htbp]
    \centering
    \includegraphics[width=\columnwidth]{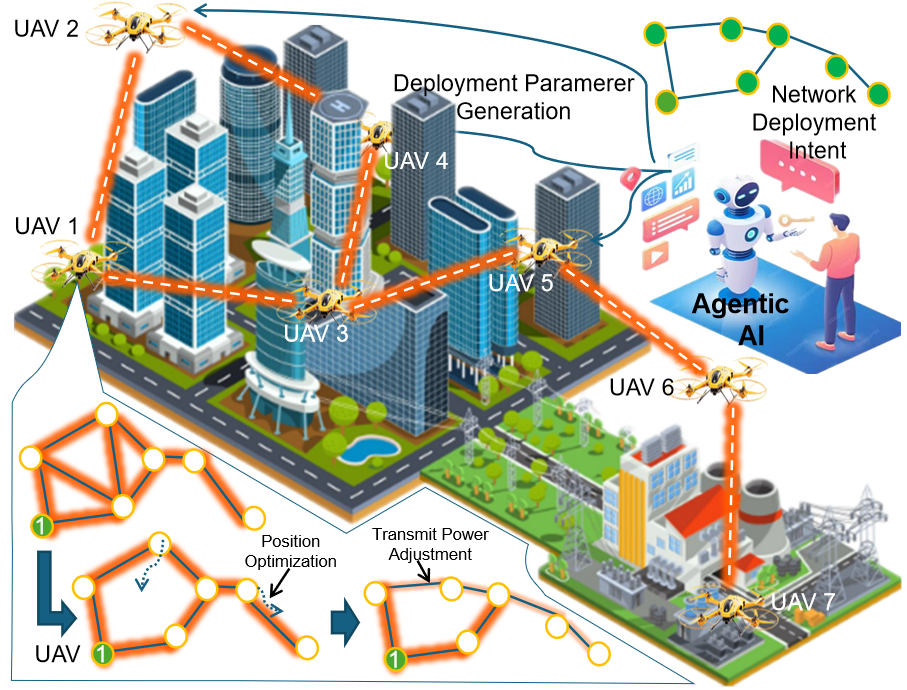} 
    \caption{Agentic AI–driven UAVN deployment model. An Agentic AI system translates UAVN deployment intents into structured parameters for an EPG. Each UAV optimizes its coordinates and transmit power in a distributed manner, leading to dynamic topology evolution. The network converges to a stable NE aligned with the specified deployment intent.}
    \label{fig:System model1}
\end{figure}

The system model is illustrated in Fig. \ref{fig:System model1}. This scenario utilizes multiple UAVs to form an aerial communication platform that provides wireless access services to ground user (GU). The proposed Agentic AI-driven framework follows an intent-driven paradigm, where a high-level deployment intent is interpreted by the Agentic AI to coordinate the interaction between the UAV and the environment. The Agentic AI translates UAV deployment intents into structured parameters for an EPG. Simultaneously, an LLM is introduced, leveraging multi-source knowledge base collaborative RAG technology to assist UAVN architects in generating the utility functions and weight coefficients required for EPG models, thereby realizing scenario-adaptive weight configuration. Each UAV acts as an agent that maintains a continuous perception-reasoning-interaction loop to enhance network adaptability. Specifically, the perception capability allows each UAV to monitor local network states, including real-time signal strength, neighbor coordinates, and GU distribution. Based on these perceived environmental semantics, the reasoning process is performed through the interaction between the UAV and the LLM-based knowledge enhancer to obtain adaptive utility weights. Furthermore, the interaction capability is realized via a broadcast mechanism, enabling UAVs to share their strategy updates and collaboratively reach a stable Nash equilibrium (NE) without centralized control. This autonomous interaction loop ensures that the UAVN can dynamically reconfigure its topology and resource allocation in response to task intents and environmental occlusions.


The design objective of this system model is to maximize network throughput while minimizing energy consumption and network latency by optimizing UAV coordinates, transmit power, and the number of served GUs, under the premise of guaranteeing network connectivity. The total network operation time $T$ is divided into several discrete time slots $\mathcal{T}=\{1, 2, \dots, t, \dots, T\}$, during which the UAV coordinates, link states, and channel conditions remain constant. The network consists of $N$ UAV nodes and $M$ GUs. The set of UAV nodes is denoted as $\mathcal{N}=\{1, 2, \dots, i, j, \dots, N\}$, and the set of GUs is $\mathcal{M}=\{1, 2, \dots, m, \dots, M\}$. The coordinates of UAV $i$ at time $t$ are $\boldsymbol{q}_i(t) = (x_i(t), y_i(t), z_i(t))$, where $x_i(t)$ and $y_i(t)$ are the horizontal coordinates, and $z_i(t)$ is the flight altitude. To ensure flight safety, the flight altitude of the UAVs is determined based on low-altitude airspace management regulations, terrain topography, and task requirements \cite{ieee2021low}, such that $z_{\min} \leq z_i(t) \leq z_{\max}, \forall i \in \mathcal{N}$, where $z_{\min}$ and $z_{\max}$ represent the minimum and maximum flight altitudes, respectively. The coordinate of GU $m$ is denoted as $\boldsymbol{q}_m(t) = (x_m(t), y_m(t), 0)$, assuming that the moving speed of GUs is typically much lower than that of the UAVs.


\subsection{Topology Model}

The topology of the UAVN is modeled as an undirected graph $G(t)=({N}, \mathcal{E}(t))$, where $\mathcal{E}(t)$ is the set of communication links at time $t$. To ensure the cooperative communication capability and information transmission reachability of the network, $G(t)$ is required to remain connected at any given time, implying that there exists at least one communication path between any two UAVs. The network topology is modeled as a symmetric adjacency matrix $\boldsymbol{A}(t)=[a_{ij}(t)]_{N \times N}$, with elements defined as follows:
\begin{equation}
    a_{ij}(t) = 
    \begin{cases} 
    1, & \text{if a communication link exists} \\
       & \text{between UAV } i \text{ and } j \\
    0, & \text{otherwise}
    \end{cases}.
\end{equation}

Since the network is an undirected graph, the adjacency matrix satisfies symmetry:
\begin{equation}
    a_{ij}(t) = a_{ji}(t), \quad a_{ii}(t) = 0.
\end{equation}

The maximum communication radius of a UAV is $R_c$, which is jointly determined by the UAV's transmit power, reception sensitivity, antenna gain, and channel propagation characteristics. In practical systems, to ensure link quality and communication reliability, constraints on the Signal-to-Noise Ratio (SNR) or Signal-to-Interference-plus-Noise Ratio (SINR) thresholds are typically considered. The Euclidean distance between UAV $i$ and $j$ at time $t$ is expressed as follows:
\begin{equation}
\begin{split}
    d_{ij}(t) &= ||\boldsymbol{q}_i(t) - \boldsymbol{q}_j(t)|| \\
    &= \sqrt{(x_i(t)\!-\!x_j(t))^2 \!+\! (y_i(t)\!-\!y_j(t))^2 \!+\! (z_i(t)\!-\!z_j(t))^2}
\end{split}.
\end{equation}

This paper defines the LoS link indicator $\zeta_{ij}(t) \in \{0, 1\}$ as follows:

\begin{equation}
\zeta_{ij}(t) = 
\begin{cases} 
1, & \parbox[t]{0.25\textwidth}{LoS,} \\
\vspace{1mm} \\ %
0, & \parbox[t]{0.25\textwidth}{NLoS.}
\end{cases}
\end{equation}

The determination of an LoS link is based on the altitudes of the UAVs, their horizontal distance, and the distribution of obstacles in the environment. Specifically, $\zeta_{ij}(t) = 1$ when the direct path between UAVs $i$ and $j$ is not obstructed by buildings, terrain, or other obstacles; otherwise, $\zeta_{ij}(t) = 0$.

Network connectivity relies on the algebraic properties of the graph's Laplacian matrix $\boldsymbol{L}(t) = \boldsymbol{\nabla}(t) - \boldsymbol{A}(t)$, where $\boldsymbol{\nabla}(t) = \text{diag}(\delta_1, \delta_2, \dots, \delta_N)$ is the degree matrix, and $\delta_i = \sum_{j \in \mathcal{N}} a_{ij}(t)$ is the degree of node $i$. A necessary and sufficient condition for network connectivity is that the second smallest eigenvalue (algebraic connectivity) of the Laplacian matrix is greater than \cite{alamdar2025decentralized}.

\begin{table}[!htbp]
    \caption{Selected Symbols and Definitions}
    \label{tab:symbols}
    \centering
    \renewcommand{\arraystretch}{1.2}
    
    \begin{tabular}{c|c|p{5.8cm}} 
        \hline
        & \textbf{Symbol} & \textbf{Definition} \\
        \hline
       
        \multirow{16}{*}{\rotatebox{90}{\textbf{System Model}}} 
        & $\mathcal{N}, N$ & Set of UAVs and the number of UAVs \\
        & $\mathcal{M}, M$ & Set of GUs and the number of users \\
        & $\mathcal{T}, T$ & Set of time slots and total operation time \\
        & $\boldsymbol{q}_i(t)$ & Coordinates vector of UAV $i$ at time $t$ \\
        & $\boldsymbol{v}_i(t)$ & Velocity vector of UAV $i$ at time $t$ \\
        & $\boldsymbol{A}(t)$ & Network adjacency matrix \\
        & $a_{ij}(t)$ & Link status between UAV $i$ and $j$ \\
        & $\boldsymbol{C}(t)$ & User association matrix \\
        & $c_{im}(t)$ & Indicator if UAV $i$ serves user $m$ \\
        & $p_i(t)$ & Transmit power of UAV $i$ \\
        & $R_c$ & Maximum communication radius \\
        & $\boldsymbol{L}(t)$ & Laplacian matrix of the network graph \\
        & $\lambda_2(\cdot)$ & Algebraic connectivity (2nd smallest eigenvalue) \\
        & $Th_{\text{total}}$ & Total network throughput \\
        & $E_{\text{total}}$ & Total energy consumption \\
        & $\mathbb{T}_{\text{total}}$ & Total network latency \\
        \hline

        \multirow{10}{*}{\rotatebox{90}{\textbf{Algorithm}}} 
        & $P$ & The formulated MINLP optimization problem \\
        & $P1, P2$ & Decomposed discrete and continuous sub-problems \\
        & $\mathscr{F}_1, \mathscr{F}_2$ & EPG models \\
        & $\rho_{i}(\cdot)$ & Local utility function of UAV $i$ \\
        & $\Phi(\cdot)$ & Global potential function \\
        & $\Gamma_i$ & Strategy space of UAV $i$ \\
        & $\eta_{1,2,3}$ & Weight coefficients for L3-EPG utility \\
        & $\psi_{1,2,3}$ & Weight coefficients for AG-EPG utility \\
        & $\mathrm{U}_{\Theta}$ & LLM-based retrieval model \\
        & $\Xi$ & Generated response sequence by LLM \\
        \hline
    \end{tabular}
\end{table}

\subsection{Communication Model}

The communication links in the UAVN are classified into U2U and U2G. U2U typically possesses LoS conditions but is still subject to path loss, small-scale fading, and doppler effects. U2G has more complex channel conditions due to the influence of building blockage, terrain undulation, and ground reflection.

\subsubsection{U2U Communication}
The path loss of the U2U distinguishes between LoS and Non-Line-of-Sight (NLoS) cases to reflect the impact of potential obstacles in aerial environments:
\begin{equation}
L_{\text{U2U}}(d_{ij}(t)) = 
    \begin{cases} 
    20\log_{10}\left(\frac{4\pi f_c d_{ij}(t)}{3 \times 10^8}\right) + \xi_{\text{LoS}}, & \zeta_{ij}(t)=1 \\
    20\log_{10}\left(\frac{4\pi f_c d_{ij}(t)}{3 \times 10^8}\right) + \xi_{\text{NLoS}}, & \zeta_{ij}(t)=0
    \end{cases},
\end{equation}
where $f_c$ denotes the carrier frequency, $\xi_{\text{LoS}}$ and $\xi_{\text{NLoS}}$ represent the additional losses for LoS and NLoS links respectively. The channel power gain from UAV $i$ to UAV $j$ is then expressed as $h_{ij} = G_{tx} G_{rx} \cdot 10^{-L_{\text{U2U}}(d_{ij}(t))/10}$, where $G_{tx}$ and $G_{rx}$ are the gains of the transmitting and receiving antennas respectively. $PL_{\mathrm{U2U}}$ is the path loss of the U2U.

When UAV $j$ receives a signal, it suffers from interference from other simultaneously transmitting UAVs $k$ ($k \neq i, k \in \mathcal{N}$). The total interference-plus-noise power is $I_j(t) = \sum_{k \neq i, k \in \mathcal{N}} p_k(t) h_{kj} + \sigma^2$, where $\sigma^2 = \kappa_B T_0 B$ is the power of additive white Gaussian noise, $\kappa_B$ is the Boltzmann constant, $T_0$ is the absolute temperature, and $B$ is the system bandwidth.

The SINR of the link from UAV $i$ to $j$ is expressed as follows:
\begin{equation}
    \gamma_{ij} = \frac{P_{ij}^{rx}}{\sum_{k \neq i, k \in \mathcal{N}} p_k(t) h_{kj} + \sigma^2},
\end{equation}
where $P_{ij}^{rx} = p_i(t)h_{ij}$ is the signal power received by UAV $j$ from UAV $i$, and $p_i$ is the transmit power of UAV $i$.

Based on the Shannon theory, the instantaneous data transmission rate of the UAV link $(i,j)$ is expressed as follows:
\begin{equation}
    r_{ij}(t) = B \log_2(1 + \gamma_{ij}).
\end{equation}

\subsubsection{U2G Communication}
According to ITU-R P.1410 recommendation \cite{itu2019propagation}, the LoS probability $P_{\text{LoS}}$ of the U2G is related to the elevation angle $\theta$:
\begin{equation}
    P_{\text{LoS}}(\theta) = \frac{1}{1 + \vartheta \exp(-\beta[\theta - \vartheta])},
\end{equation}
where $\theta = \arctan\left(\frac{z_i(t)}{\sqrt{(x_i(t)-x_m(t))^2 + (y_i(t)-y_m(t))^2}}\right)$ is the angle between the GU's horizontal plane and the U2G, and $\vartheta, \beta$ are environmental parameters.

The path loss of the U2G distinguishes between LoS and NLoS cases:
\begin{equation}
    L_{\text{U2G}}(d_{im}(t)) = 
    \begin{cases} 
    20\log_{10}\left(\frac{4\pi f_c d_{im}(t)}{3 \times 10^8}\right) + \xi_{\text{LoS}}, & \zeta_{im}(t)=1 \\
    20\log_{10}\left(\frac{4\pi f_c d_{im}(t)}{3 \times 10^8}\right) + \xi_{\text{NLoS}}, & \zeta_{im}(t)=0
    \end{cases},
\end{equation}
where $d_{im}(t) = ||\boldsymbol{q}_i(t) - \boldsymbol{q}_m(t)||$ represents the distance between UAV $i$ and user $m$, and $\xi_{\text{NLoS}}$ is the additional NLoS loss.

Considering the interference caused by multiple UAVs transmitting simultaneously, the SINR for user $m$ is:
\begin{equation}
    \gamma_{im} = \frac{P_{im}^{rx}}{\sum_{k \neq i, k \in \mathcal{N}} p_k(t) h_{km} + \sigma^2},
\end{equation}
where $h_{km}$ is the channel power gain from UAV $k$ to user $m$, denoted as $h_{km} = G_{tx} G_{rx} \cdot 10^{-PL_{\text{U2G}}(d_{km}(t))/10} \cdot |\vartheta|^2$. Here, $\vartheta$ is the small-scale fading coefficient, determined by $P_{\text{LoS}}$. $PL_{\mathrm{U2G}}$ is the path loss of the U2G. $P_{im}^{rx} = p_i(t)h_{im}$ is the signal power received by user $m$ from UAV $i$.

The data transmission rate of the U2G is:
\begin{equation}
    r_{im}(t) = B \log_2(1 + \gamma_{im}).
\end{equation}

\subsubsection{Throughput}
The total throughput of U2U is:
\begin{equation}
    Th_{\text{U2U}}(t) = \sum_{i \in \mathcal{N}} \sum_{j > i, j \in \mathcal{N}} a_{ij}(t) r_{ij}(t).
\end{equation}

The total throughput of U2G is:
\begin{equation}
    Th_{\text{U2G}}(t) = \sum_{i \in \mathcal{N}} \sum_{m \in \mathcal{M}} c_{im}(t) r_{im}(t),
\end{equation}
where $c_{im}(t)$ is the coverage indicator variable. If GU $m$ communicates with UAV $i$, then $c_{im}(t)=1$; otherwise, $c_{im}(t)=0$.
The throughput of UAV $i$ at time $t$ is $Th_i(t)$, and the network throughput is the sum of U2U and U2G throughputs:
\begin{equation}
    Th_{\text{total}}(t) = \sum_{i \in \mathcal{N}} Th_i(t) = Th_{\text{U2U}}(t) + Th_{\text{U2G}}(t).
\end{equation}

\subsection{Mobility Model}
The mobility characteristics of UAVs directly affect network topology dynamics and link stability. Under the discrete time slot model, the UAV coordinates update equation is:
\begin{equation}
    \boldsymbol{q}_i(t+1) = \boldsymbol{q}_i(t) + \boldsymbol{v}_i(t) \Delta t.
\end{equation}

Expanding this vector equation into three coordinate components:
\begin{equation}
    \begin{cases}
    x_i(t+1) = x_i(t) + v_i^x(t) \Delta t \\
    y_i(t+1) = y_i(t) + v_i^y(t) \Delta t \\
    z_i(t+1) = z_i(t) + v_i^z(t) \Delta t
    \end{cases}.
\end{equation}

The UAV velocity is controlled by acceleration:
\begin{equation}
    v_i(t+1) = v_i(t) + \dot{\boldsymbol{v}}_i(t) \Delta t,
\end{equation}
where $\dot{\boldsymbol{v}}_i(t)$ is the acceleration vector.

\subsection{Energy Consumption Model}
The total energy consumption of a UAV comprises communication energy consumption and flight energy consumption. These are critical factors constraining network performance and task duration.

\subsubsection{Communication Energy Consumption}
The energy consumption of UAV $i$ includes transmission energy and circuit energy:
\begin{equation}
    E_{\text{com},i}(t) = p_i(t) t_{\text{com}} + P_{\text{circ}}(t) t_{\text{com}},
\end{equation}
where $p_i(t)$ is the transmit power, and $t_{\text{com}}$ represents the communication duration, with length $\Delta t$. $P_{\text{circ}}(t)$ is the radio frequency circuit power, including the power consumption of amplifiers, mixers, filters, Analog to Digital Converter, etc. The communication energy of UAV $i$ simplifies to:
\begin{equation}
    E_{\text{comm},i}(t) = (p_i(t) + P_{\text{circ}}(t)) \Delta t.
\end{equation}

\subsubsection{Flight Energy Consumption}
The energy consumption consists of blade profile power, induced power, parasite power, and acceleration energy. 
The blade profile power is used to overcome air resistance during rotor blade rotation: $P_{\text{blade}}(t) = \frac{\delta}{8} \rho_{\text{air}} S \kappa_{\text{tip}}^3$, where $\delta$ is the profile drag coefficient, $\rho_{\text{air}}$ is the air density, $S$ is the rotor disk area, and $\kappa_{\text{tip}}$ is the blade tip speed.
Induced power is used to generate lift to support the UAV weight: $P_{\text{induced}}(t) = \frac{(1+\kappa_b)W^{3/2}}{\sqrt{2\rho_{\text{air}}S}}$, where $W$ is the UAV weight (frame, battery, payload), and $\kappa_b$ is the induced power correction factor. This power is maximum at hover and decreases slightly with increased flight speed.
Parasite power accounts for air resistance on the fuselage during forward flight: $P_{\text{parasite}}(t) = \frac{1}{2} d_f v_i(t)^3$, where $d_f$ is the fuselage drag coefficient, and $v_i(t)$ is the flight speed of the UAV $i$.

The total flight power of UAV $i$ during flight is $P_{\text{flight},i}(t) = P_{\text{blade}}(t) + P_{\text{induced}}(t) + P_{\text{parasite}}(t)$. Within $\Delta t$, the flight energy consumption is $E_{\text{flight},i}(t) = P_{\text{flight},i} \Delta t$. When the UAV accelerates or decelerates, additional energy is required to change its kinetic energy: $E_{\text{accel},i}(t) = \frac{1}{2}(|v_i(t+\Delta t)|^2 - |v_i(t)|^2)W/g$, where $g$ is the gravitational acceleration. The total flight energy is:
\begin{equation}
    E_{\text{flight},i}(t) = P_{\text{flight},i} \Delta t + E_{\text{accel},i}(t).
\end{equation}

The energy consumption of UAV $i$ at time $t$ is $E_i(t)$, and the energy consumption of all UAVs in the network is:
\begin{equation}
    E_{\text{total}}(t)  = \sum_{i \in \mathcal{N}} E_i(t) = \sum_{i \in \mathcal{N}} E_{\text{comm},i}(t) + E_{\text{flight},i}(t) .
\end{equation}

\subsection{Latency Model}
Transmission latency is the time required to send a packet over the link, depending on the packet size and transmission rate.
\begin{equation}
    T_{\text{trans},ij}(t) = \frac{L_{\text{data}}}{r_{ij}(t)},
\end{equation}
where $L_{\text{data}}$ is the packet size. Propagation latency is the time required for the signal to travel through space:
\begin{equation}
    T_{\text{prop},ij}(t) = \frac{d_{ij}(t)}{3 \times 10^8}.
\end{equation}

The average latency for UAV $i$ is $\mathbb{T}_i(t)$, and the network latency is the sum of the latencies for all active communication links:
\begin{equation}
\begin{aligned}
\mathbb{T}_{\text{total}} = & \sum_{i\in\mathcal{N}} \mathbb{T}_i(t) \\
= & \sum_{i\in\mathcal{N}} \sum_{j>i,j\in\mathcal{N}} a_{ij}(t) (T_{\text{trans},ij}(t) + T_{\text{prop},ij}(t)) \\
& + \sum_{i\in\mathcal{N}} \sum_{m\in\mathcal{M}} c_{im}(t) (T_{\text{trans},im}(t) + T_{\text{prop},im}(t)),
\end{aligned}
\label{eq:21}
\end{equation}
where the first double summation term represents the cumulative latency of the U2U links, where $a_{ij}(t)$ is the network adjacency indicator. The second term denotes the total access latency between U2G links, where $c_{im}(t)$ is the coverage indicator. Specifically, $T_{\text{trans},im}(t) = L_{\text{data}} / r_{im}(t)$ and $T_{\text{prop},im}(t) = d_{im}(t) / (3 \times 10^8)$ represent the transmission delay and propagation delay for user $m$ associated with UAV $i$, respectively.

\begin{figure*}[htbp]
    \centering
    \includegraphics[width=\textwidth]{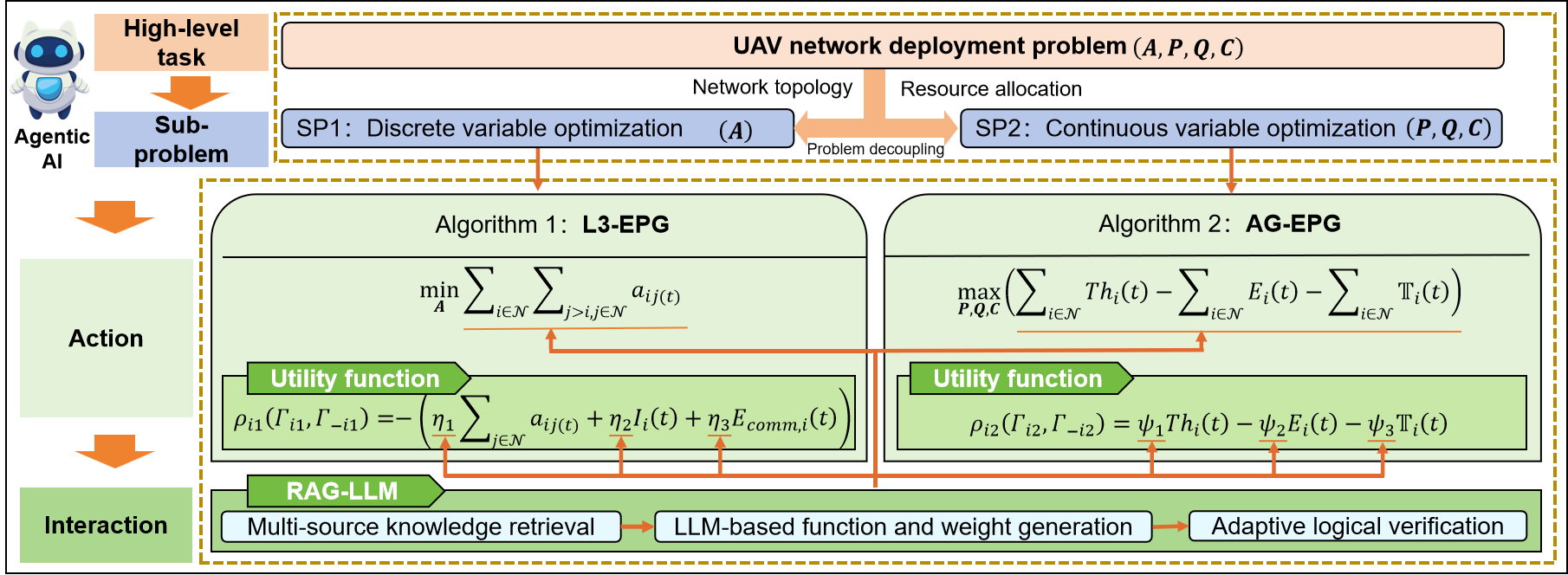}
    
    \caption{The decomposition process of the Agentic AI-driven UAVN deployment optimization problem. Agentic AI is capable of decoupling high-level tasks, generate sub-problems, plan actions, and interact with other tools or environments with human \cite{zhang2026toward}. First, the network deployment problem is divided into two subproblems. Then, these subproblems are solved using L3-EPG and AG-EPG, respectively. Finally, the RAG-LLM module is used to generate adaptive weight coefficients for the utility function through multi-source knowledge retrieval and logical verification, aiming to obtain a network deployment scheme and parameters that meet the actual needs of the network.}
    \label{fig:framework}
\end{figure*}

\section{Problem Formulation}
\label{sec:4}
This section mathematically formulates the UAVN deployment task as an MINLP. First, we define the comprehensive objective function, which encompasses network topology, throughput, energy, and latency. Subsequently, we introduce the physical system constraints, paving the way for the specific problem-solving strategies detailed in the following sections. The optimization objectives for the UAVN are to minimize the number of communication links between UAVs, maximize network throughput, and minimize UAV energy consumption and network latency:

\begin{equation}
\label{eq:problem_P} 
    \begin{split}
        P: \min_{\boldsymbol{A}, \mathbf{P}, \mathbf{Q}, \mathbf{C}} \bigg( & \sum_{i \in \mathcal{N}} \sum_{j>i, j \in \mathcal{N}} a_{ij}(t) \\
        & - \Upsilon Th_{\text{total}} + \mu E_{\text{total}} + \tau \mathbb{T}_{\text{total}} \bigg)
    \end{split}
\end{equation}

\addtocounter{equation}{-1}
\begin{subequations}
\label{eq:constraints_all} 
\begin{align}
    \text{s.t.} \quad & \sum_{i \in \mathcal{N}} c_{im}(t) \geq 1, \quad \forall m \in \mathcal{M}, \label{eq:24a} \\
    & d_{im}(t) \leq R_c, \quad \text{if } c_{im}(t)=1, \forall i \in \mathcal{N}, m \in \mathcal{M}, \label{eq:24b} \\
    & p_{\min} \leq p_i(t) \leq p_{\max}, \quad \forall i \in \mathcal{N}, \label{eq:24c} \\
    & \boldsymbol{q}_i(t) \in \mathbb{R}^3, \quad \forall i \in \mathcal{N}, \label{eq:24d} \\
    & \lambda_2(\boldsymbol{L}(t)) > 0, \label{eq:24e} \\
    & E_i(t) \leq E_i^{\max}, \label{eq:24f}
\end{align}
\end{subequations}
where $\boldsymbol{A}$ is the adjacency matrix, $\mathbf{P}$ is the power allocation vector, $\mathbf{Q}$ is the coordinates vector, and $\mathbf{C}$ is the user association matrix. $\Upsilon$, $\mu$, and $\tau$ are weighting factors. $R_c$ is the coverage radius, and $p_{\min}$ and $p_{\max}$ are the minimum and maximum transmit powers, respectively. $\mathbb{R}^3$ denotes the 3D region of the UAVs. \eqref{eq:24a} represents the user coverage constraint, ensuring that every GU $m$ is covered by at least one UAV $i$ providing communication service. \eqref{eq:24b} represents the communication distance constraint; when user $m$ establishes a connection with UAV $i$, the Euclidean distance $d_{im}(t)$ must not exceed the maximum communication radius $R_c$. \eqref{eq:24c} represents the UAV transmit power constraint, where $p_{\min}$ and $p_{\max}$ denote the minimum and maximum allowable transmit powers, respectively. \eqref{eq:24d} specifies the spatial coordinate constraints for the UAV, requiring that the coordinates $\mathbf{q}_i(t)$ of UAV $i$ must be located within the permissible flight region $\mathbb{R}^3$. \eqref{eq:24e} is the network connectivity constraint, ensuring that the network graph $G(t)$ remains connected at all times by requiring the second smallest eigenvalue $\lambda_2$ of the Laplacian matrix $\boldsymbol{L}(t)$ to be greater than zero. \eqref{eq:24f} indicates the upper bound of energy consumption for UAV $i$, where $E_i^{\max}$ is the initial battery energy of UAV $i$. 

Since the problem of UAVN deployment involves both discrete and continuous variables and contains non-convex constraints, $P$ is an MINLP. It belongs to the class of NP-hard problems with variables coupled to each other. The strategy space grows exponentially with the network scale, making centralized solving computationally prohibitive and ill-suited for dynamic environments.

To address this, we leveraged the Agentic AI paradigm to orchestrate a logical problem decomposition as illustrated in Fig. \ref{fig:framework}. In this framework, Agentic AI served as an intelligent task planner that translated the deployment intent into decentralized and tractable sub-problems. The Agentic AI partitioned the complex task into a discrete optimization layer denoted as SP1 and a continuous optimization layer denoted as SP2. The action block in the figure represented the execution of game-theoretic learning dynamics. Specifically, Agentic AI strategically decoupled $P$ into two sequential and cooperative sub-problems: $P1$ and $P2$.

1) Discrete Topology Optimization: At the large spatial scale, the macroscopic structure of the UAVN is defined by discrete link decisions. As depicted in the SP1, Agentic AI identifies that the adjacency matrix $\boldsymbol{A}$ dictates the basic connectivity and interference levels. To fulfill the intent of maintaining a sparse yet connected backbone, $P1$ is decoupled to minimize the number of communication links:
\begin{equation}
    P1: \min_{\boldsymbol{A}} \sum_{i \in \mathcal{N}} \sum_{j>i, j \in \mathcal{N}} a_{ij}(t).
\end{equation}

2) Continuous Resource and Deployment Optimization: Following the macroscopic topology configuration, Agentic AI directs individual agents to fine-tune their states at a small spatial scale as shown the SP2. Since network throughput, UAV energy consumption, and network latency are continuous variables, $P2$ is formulated as a continuous optimization to maximize the system comprehensive utility:
\begin{equation}
    P2: \max_{\mathbf{P}, \mathbf{Q}, \mathbf{C}} \left( \sum_{i \in \mathcal{N}} Th_i(t) - \sum_{i \in \mathcal{N}} E_i(t) - \sum_{i \in \mathcal{N}} \mathbb{T}_i(t) \right).
\end{equation}

Through this Agentic AI-driven decomposition, the intractable MINLP is transformed into a dual-scale game-theoretic learning process. The specific impact of the Agentic AI paradigm manifests in three key ways that address the previously mentioned drawbacks of traditional methods. First, it mitigates the computational and centralized bottlenecks identified in \textbf{Challenge 1} by delegating high-dimensional decision-making to autonomous UAVs. This distributed approach significantly reduces the overhead and eliminates single-point failures. Second, to overcome the instability of heuristic methods in \textbf{Challenge 2}, Agentic AI employs game-based learning with a consistent mapping to a global potential function, providing a rigorous mathematical guarantee for convergence to a NE. Finally, addressing the generalization and manual-tuning limitations in \textbf{Challenge 3}, the Agentic AI incorporates an LLM-based knowledge enhancer. This allows the system to autonomously formulate optimal utility weights by reasoning through diverse scenarios, thereby alleviating the heavy reliance on expert-in-the-loop parameterization and enhancing the adaptability of UAVN in heterogeneous environments.

\section{UAVN Deployment based on EPG}
\label{sec:5}
In this section, we develop a dual spatial-scale optimization framework based on EPG. Specifically, Section\ref{sec:5}-A introduces a L3-EPG algorithm to optimize discrete network topology configurations at a large spatial scale. Subsequently, Section\ref{sec:5}-B presents an AG-EPG algorithm to jointly optimize continuous variables, including global throughput, UAV energy consumption, and network latency at a small spatial scale.

$P1$ and $P2$ are characterized as multi-participant optimization problems. Due to constraints such as restricted scenarios and a large number of participants, the use of centralized solution methods is associated with non-negligible latency, high computational volume, and high complexity \cite{sun2024competitive}. Consequently, these problems are deemed more suitable for resolution via distributed methods \cite{hu2023multi}. Previous work has demonstrated that potential games serve as powerful tools for solving distributed problems \cite{monderer1996potential}. In a potential game, the change in the utility function is consistent with the change in the global potential function, thereby guaranteeing the convergence and global consistency of the game \cite{poveda2023fixed}. In particular, EPGs can strictly guarantee the existence of a Pure Strategy NE (PSNE) \cite{zhang2024decentralized}. In this paper, the method of EPGs is utilized to solve $P1$ and $P2$. The topology optimization within the UAVN is modeled as a non-zero-sum, non-cooperative game. $P1$ is modeled as an EPG model $\mathscr{F}_{1}$, which is utilized to minimize communication links between UAVs and optimize discrete variables. $P2$ is modeled as an EPG model $\mathscr{F}_{2}$, which is utilized to optimize global throughput, UAV energy consumption, and network latency by optimizing continuous variables. Utility functions and potential functions satisfying the conditions of EPGs are designed for $\mathscr{F}_{1}$ and $\mathscr{F}_{2}$. This ensures that the change in the utility function caused by a unilateral strategy change of any UAV is consistent with the change in the potential function. The game achieves a global optimal solution through the NE, ensuring the convergence of the distributed optimization.
\renewcommand{\algorithmicrequire}{\textbf{Input:}}
\renewcommand{\algorithmicensure}{\textbf{Output:}}

\begin{algorithm}[!htbp]
    \caption{L3-EPG} 
    \label{alg:l3_epg}
    
    \begin{algorithmic}[1]
        \Require $N$, $R_c$, $T_{\max}$
        \Ensure $\boldsymbol{A}^*$
        
        \State Generate initial coordinates based on K-Means, construct $G(0)$ and $\boldsymbol{A}(0)$
        \State Initialize: $a_{ij}(0) \gets 1, \forall i \in \mathcal{N}$; $t \gets 0$
        
        \While{$\|\sum a_{ij}(t) - \sum a_{ij}(t-1)\| \geq 1$ and $t < T_{\max}$}
            
            \For{each $i \in \mathcal{N}$}
                \State Calculate $\rho_{i1}(\Gamma_{i1}, \Gamma_{-i1})$
                
                \If{$a_{ij}(t)=1 \ \& \ \zeta_{ij}(t)=1$}
                    \State Update $\Gamma_{i1}(t+1)$ with probability $\exp[\rho_{i1}(\Gamma_{i1}, \Gamma_{-i1})] / \sum \exp[\rho_{i1}]$
                    \State $a_{ij}(t+1) \gets 0$
                \EndIf
            \EndFor
            
            \State Construct candidate adjacency matrix $\boldsymbol{A}'(t+1)$
            
            \If{$\lambda_2(\boldsymbol{L}(t)) > 0, \ \text{isConnected}(\boldsymbol{A}'(t+1))$}
                \State $\boldsymbol{A}(t+1) \gets \boldsymbol{A}'(t+1)$
            \Else
                \State $\boldsymbol{A}(t+1) \gets \boldsymbol{A}(t)$
            \EndIf
            
            \State $t \gets t+1$
        \EndWhile
        
        \State \Return $\boldsymbol{A}^* \gets \boldsymbol{A}(t)$
    \end{algorithmic}
\end{algorithm}

\subsection{Log-Linear Learning based EPG}
To address $P1$, a L3-EPG is proposed, as shown in Algorithm \ref{alg:l3_epg}. This algorithm is utilized to optimize the number of communication links between UAVs. The L3 method is based on DRL principles and adopts a probabilistic strategy selection mechanism. In this mechanism, each UAV acts as an independent participant and autonomously decides whether to maintain or disconnect a communication link based on local information encoded in the adjacency matrix. This decision process is guided by log-linear learning rules, balancing exploration and exploitation by allocating action probabilities proportional to utility. This mechanism ensures that UAVs adaptively optimize their strategies during the iteration process, minimizing redundant links while maintaining necessary network connectivity. The EPG model is defined as $\mathscr{F}_{1}=\{N,\{\Gamma_{i1}\}_{i\in\mathcal{N}},\{\rho_{i1}\}_{i\in\mathcal{N}}\}$, where $\rho_{i1}$ represents the utility of UAV $i$. $\Gamma_i^1=\{ a_{ij} ( t ) , i , j \in \mathcal{N} \}$ represents the strategy space of the UAV, meaning the strategy of each UAV is to adjust the communication link $a_{ij}$ with other UAVs.

Each UAV enables the UAVN to achieve an optimal connection topology by adjusting communication links between UAVs and maximizing its local utility. A utility function is designed to reflect the reduction of redundant communication links between UAVs while simultaneously lowering interference and communication energy consumption. The utility function for UAV $i$ is defined as:
\begin{equation}
\begin{split}
\label{eq:28}
    \rho_{i1}(\Gamma_{i1},\Gamma_{-i1})&=-(\eta_1\sum_{j\in\mathcal{N}}a_{ij}(t)+ \\
    &\eta_2I_i(t)+\eta_3E_{comm,i}(t)),\\
\end{split}
\end{equation}
where $\Gamma_{-i1}$ represents the strategies of other UAV, $\sum_{j\in\mathcal{N}}a_{ij}(t)$ is the number of links connected to UAV $i$, and $\eta_1, \eta_2, \eta_3$ are weights.

The potential function can directly represent the changes in the utility functions of the game participants, ensuring that the utility of any participant is consistent with the global objective. To align individual decisions with the global optimal network topology, the global potential function is defined as:
\begin{equation}
\Phi_1(\Gamma)=\sum_{i\in \mathcal{N} }\rho_{i1}(\Gamma_{i1},\Gamma_{-i1}).
\end{equation}

Furthermore, based on the iterative update mechanism \cite{zhao2022multi} of BRD, each UAV calculates and selects the best action that enhances its local utility after receiving strategy information from other UAVs:
\begin{equation}
\label{eq:30}
\Gamma_{i1}(t+1){=}\mathrm{argmax}\rho_{i1}(\Gamma_{i1},\Gamma_{-i1}(t)).
\end{equation}

L3-EPG addresses the link selection problem involving discrete parameters, and its optimization process focuses primarily on global optimality at the link topology level. Although occlusion factors have been considered in the utility function and links less affected by occlusion are prioritized, individual links may still be partially obstructed by buildings under certain initial UAV coordinates. This issue will be resolved in the subsequent continuous parameter optimization phase of the algorithm \ref{alg:ag_epg}.

\subsection{Approximate Gradient based EPG algorithm}
To address $P2$, an AG-EPG algorithm is proposed to enhance the effectiveness of resource allocation, as shown in Algorithm \ref{alg:ag_epg}. AG-EPG enables each UAV to iteratively optimize its decision space via gradient-based best responses and historical action results, identifying optimal actions consistent with network objectives derived from the strategies of other UAVs. This method includes a random exploration process wherein UAVs utilize updated condition evaluation strategies through an broadcast mechanism, ensuring necessary real-time data sharing. The AG-EPG algorithm performs continuous parameter optimization based on the optimization completed by the L3-EPG algorithm. Unlike BRD, which selects the optimal action solely based on the current strategies of other participants, the proposed AG-EPG algorithm combines approximate gradient information with historical decision results for best response calculation and introduces a random exploration mechanism to avoid falling into local optima, thereby enhancing the algorithm's global search capability. The EPG model is defined as $\mathscr{F}_{2}=\{N,\{\Gamma_{i2}\}_{i\in\mathcal{N}},\{\rho_{i2}\}_{i\in\mathcal{N}}\}$, where $\Gamma_{i2}=\{\mathbf{q}_{i}(t),p_{i}(t),\mathbf{c}_{i}(t)\}$ represents the strategy space of the UAVs. The strategy of each UAV involves adjusting its coordinates $\mathbf{q}_{i}(t)$, transmission power $p_i ( t )$, and the GU equipment $\mathbf{c}_i(t){=}\{c_{im}(t),m{\in}\mathcal{M}\}$ with which it communicates. The utility function for UAV $i$ is defined as:
\begin{equation}
\label{eq:31}
\rho_{i2}(\Gamma_{i2},\Gamma_{-i2})=\psi_1Th_i(t)-\psi_2E_i(t)-\psi_3\mathbb{T}_{i}(t)\,
\end{equation}
where $\Gamma_{-i2}$ represents the strategies of other UAVs, and $\psi_1, \psi_2, \psi_3$ are weight coefficients. All participants aim to maximize their own utility, i.e., $max\rho_{i2}(\Gamma_{i2},\Gamma_{-i2})$.

Since the utility function encourages UAVs to improve global throughput, UAVs will adjust their flight altitude to reduce the communication distance to GUs, provided that UAV links are not altered. Therefore, fine-tuning of the UAVs' coordinates is required to ensure comprehensive coverage of GUs by the UAVs.

The global potential function is defined as:
\begin{equation}
\Phi_2(\Gamma)=\sum_{i\in\mathcal{N}}\rho_{i2}\left(\Gamma_{i2},\Gamma_{-i2}\right).
\end{equation}

Furthermore, each UAV updates its link strategy \cite{chen2021coalition} according to the probability update mechanism of L3:
\begin{equation}
\label{eq:33}
\Gamma_{i2}(t+1){=}\mathrm{argmax}\rho_{i2}(\Gamma_{i2},\Gamma_{-i2}(t)).
\end{equation}

\subsection{NE Analysis and Proof}

To ensure that the strategy update mechanisms described in \eqref{eq:30} and \eqref{eq:33} lead to a stable system state, it is essential to verify the existence of a PSNE. The convergence stability of our proposed distributed optimization framework depends on whether the utility functions formulated in \eqref{eq:28} and \eqref{eq:31} satisfy the properties of an EPG. In this subsection, we provide a rigorous analytical proof to demonstrate that a unilateral strategy change by any UAV results in an identical change in the corresponding global potential function for both game models. Specifically, we first analyze game model $\mathscr{F}_{1}$, where the change in the utility function caused by a unilateral action change of each UAV is derived as follows:
\begin{equation}
\begin{split}
    & \rho_{i1}(\Gamma'_{i1}, \Gamma_{-i1}) - \rho_{i1}(\Gamma_{i1}, \Gamma_{-i1}) = \Delta \rho_{i1} \\
    & = -\eta_1 \sum_{j \in \mathcal{N}} a_{ij}(t)' - \eta_2 I_i(t)' - \eta_3 E_{comm,i}(t)' \\
    & \quad + \left( \eta_1 \sum_{j \in \mathcal{N}} a_{ij}(t) + \eta_2 I_i(t) + \eta_3 E_{comm,i}(t) \right) \\
    & = \eta_1 \sum_{j \in \mathcal{N}} (a_{ij}(t) - a_{ij}(t)') \\
    & \quad + \eta_2 (I_i(t) - I_i(t)') \\
    & \quad + \eta_3 (E_{comm,i}(t) - E_{comm,i}(t)')
\end{split}
\end{equation}

The change in the potential function caused by a unilateral action change of each UAV is:
\begin{equation}
\begin{split}
   & \Phi_{1}\left(\Gamma_{i 1}^{\prime}, \Gamma_{-i 1}\right)-\Phi_{1}\left(\Gamma_{i 1}, \Gamma_{-i 1}\right)=\Delta \Phi_{1} \\
    &=\left[\rho_{i 1}\left(\Gamma_{i 1}^{\prime}, \Gamma_{-i 1}\right)+\sum_{i \in \mathcal{N}} \rho_{i 1}\left(\Gamma_{i 1}, \Gamma_{-i 1}\right)\right]\\
    &-\left[\rho_{i 1}\left(\Gamma_{i 1}, \Gamma_{-i 1}\right)+\sum_{i \in \mathcal{N}} \rho_{i 1}\left(\Gamma_{i 1}, \Gamma_{-i 1}\right)\right]\\
    &=\rho_{i 1}\left(\Gamma_{i 1}^{\prime}, \Gamma_{-i 1}\right)-\rho_{i 1}\left(\Gamma_{i 1}, \Gamma_{-i 1}\right)\\
    &=\eta_{1} \sum_{j \in \mathcal{N}}\left(a_{i j}(t)-a_{i j}(t)^{\prime}\right)+\eta_{2}\left(I_{i}(t)-I_{i}(t)^{\prime}\right)\\
    &+\eta_{3}\left(E_{\text {comm }, i}(t)-E_{\text {comm }, i}(t)^{\prime}\right),
\end{split}
\end{equation}

When only the strategy of UAV $i$ changes from $\Gamma_{i1}$ to $\Gamma_{i1}^{\prime}$, the utilities of other UAVs remain unchanged. The change in the potential function is derived solely from the change in $\rho_{i1}$. Since $\Gamma_{-i1}$ is unchanged, only the terms of $\rho_{i1}$ change:

\begin{equation}
\Delta\Phi_1=\rho_{i1}\left(\Gamma_{i1}^{^{\prime}},\Gamma_{-i1}\right)-\rho_{i1}(\Gamma_{i1},\Gamma_{-i1})=\Delta\rho_{i1}.
\end{equation}

For game model $\mathscr{F}_{2}$, the change in the utility function caused by a unilateral action change of each UAV is:
\begin{equation}
\begin{split}
   &\rho_{i 2}\left(\Gamma_{i 2}^{\prime}, \Gamma_{-i 2}\right)-\rho_{i 2}\left(\Gamma_{i 2}, \Gamma_{-i 2}\right)=\Delta \rho_{i 2}\\
    &=\left(\psi_{1} T h_{i}(t)^{\prime}-\psi_{2} E_{i}\left(t^{\prime}\right)-\psi_{3} \mathbb{T}_{i}(t)^{\prime}\right)\\
    &-\left(\psi_{1} T h_{i}(t)-\psi_{2} E_{i}(t)-\psi_{3} \mathbb{T}_{i}(t)\right)\\
    &=\rho_{i 2}\left(\Gamma_{i 2}^{\prime}, \Gamma_{-i 2}\right)-\rho_{i 2}\left(\Gamma_{i 2}, \Gamma_{-i 2}\right)\\
    &=\psi_{1}\left(T h_{i}(t)^{\prime}-T h_{i}(t)\right)+\psi_{2}(E_{i}(t)\\
    &-E_{i}\left(t^{\prime}\right))+\psi_{3}\left(\mathbb{T}_{i}(t)-\mathbb{T}_{i}(t)^{\prime}\right).
\end{split}
\end{equation}

\begin{figure*}[htbp]
    \centering
    \includegraphics[width=\textwidth]{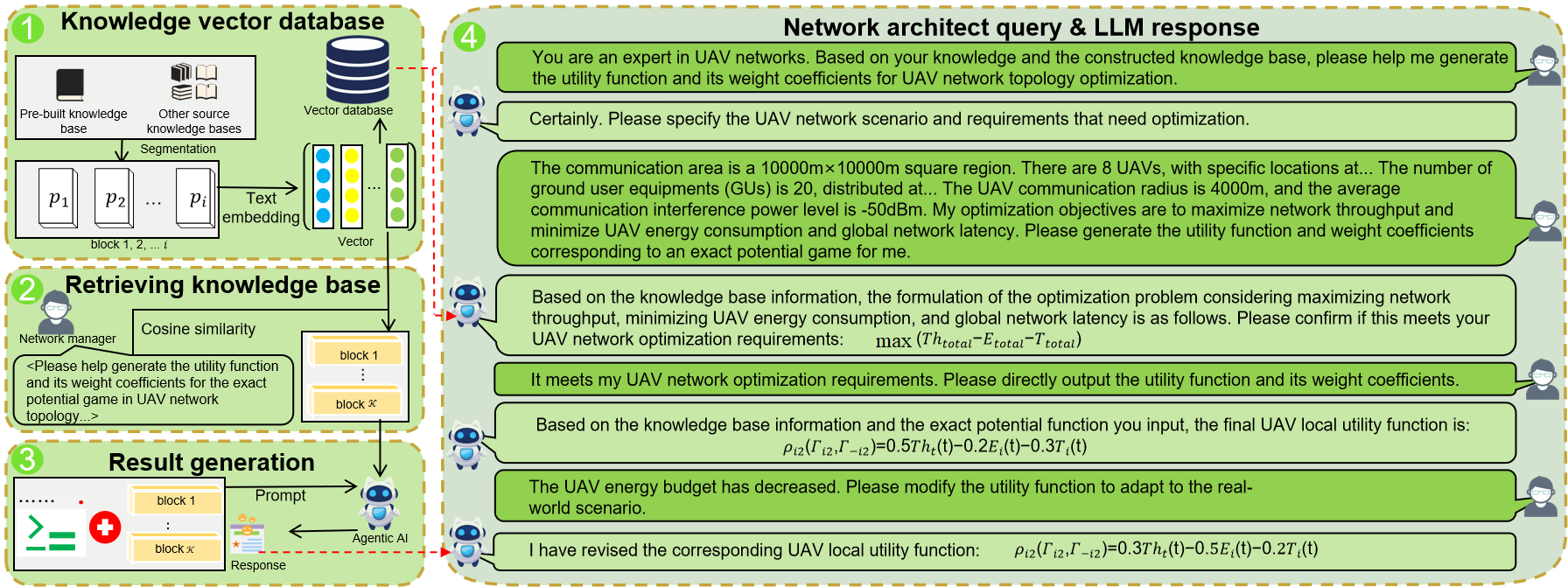}
    \caption{Framework of RAG-based LLM assisted UAVN deployment optimization. The framework consists of four modules: \textcircled{1} Knowledge Vector Database Module, which performs text segmentation and vector embedding on the pre-constructed knowledge base and literature to establish a retrievable database; \textcircled{2} Knowledge Retrieval Module, which receives the network manager's intent and retrieves the most relevant knowledge fragments via cosine similarity calculation; \textcircled{3} Result Generation Module, which feeds the retrieved domain knowledge into the LLM as contextual prompts; and \textcircled{4} Iterative Refinement Module, where the LLM generates initial utility functions and weighting coefficients. The manager provides feedback for the LLM to refine the results by integrating knowledge base information until the generated parameters satisfy the specific network requirements.}
    \label{fig:rag_framework}
\end{figure*}

The change in the potential function caused by a unilateral action change of each UAV is:
\begin{equation}
\begin{split}
   &\Phi_{2}\left(\Gamma_{i 2}^{\prime}, \Gamma_{-i 2}\right)-\Phi_{2}\left(\Gamma_{i 2}, \Gamma_{-i 2}\right)=\Delta \Phi_{2}\\
    &=\left[\rho_{i 2}\left(\Gamma_{i 2}^{\prime}, \Gamma_{-i 2}\right)+\sum_{i \in \mathcal{N}} \rho_{i 2}\left(\Gamma_{i 2}, \Gamma_{-i 2}\right)\right]\\
    &-\left[\rho_{i 2}\left(\Gamma_{i 2}, \Gamma_{-i 2}\right)+\sum_{i \in \mathcal{N}} \rho_{i 2}\left(\Gamma_{i 2}, \Gamma_{-i 2}\right)\right]\\
    &=\rho_{i 2}\left(\Gamma_{i 2}^{\prime}, \Gamma_{-i 2}\right)-\rho_{i 2}\left(\Gamma_{i 2}, \Gamma_{-i 2}\right)\\
    &=\psi_{1}\left(T h_{i}(t)^{\prime}-T h_{i}(t)\right)+\psi_{2}\left(E_{i}(t)-E_{i}(t)^{\prime}\right)\\
    &+\psi_{3}\left(\mathbb{T}_{i}(t)-\mathbb{T}_{i}(t)^{\prime}\right).
\end{split}
\end{equation}

\begin{algorithm}[!htbp]
    \caption{AG-EPG}
    \label{alg:ag_epg}
    \begin{algorithmic}[1]
        
        \Require $\mathcal{N}$, $\boldsymbol{A}^*$ from L3-EPG, $[p_{\min}, p_{\max}]$
        \Ensure $\mathbf{Q}^*$, $\mathbf{P}^*$, $\mathbf{C}^*$
        
        \State Initialize: $\boldsymbol{A}(0) \leftarrow \boldsymbol{A}^*, \Phi_2(0)$; broadcast initial state
        \State $t \leftarrow 0$; $\textit{converged} \leftarrow \text{false}$
        
        \While{\textbf{not} $\textit{converged}$}
            \For{each $i \in \mathcal{N}$ \textbf{parallel}}
                \State $\mathbf{q}_i(t)^* \leftarrow \operatorname{argmax}_{\rho_{i2}}(\mathbf{q}_i(t), \Gamma_{-i2},\zeta_{ij}(t)=1)$
                \State $p_i(t)^* \leftarrow \operatorname{argmax}_{\rho_{i2}}(p_i(t), \Gamma_{-i2},\zeta_{ij}(t)=1)$
                \State $\mathbf{c}_i(t)^* \leftarrow \operatorname{argmax}_{\rho_{i2}}(\mathbf{c}_i(t), \Gamma_{-i2},\zeta_{ij}(t)=1)$
                \State $\Gamma_{i2}(t+1) \leftarrow \{ \mathbf{q}_i(t)^*, p_i(t)^*, \mathbf{c}_i(t)^* \}$
            \EndFor
            
            \State Broadcast $\{ \mathbf{q}_i(t)^*, p_i(t)^*, \mathbf{c}_i(t)^* \}, \forall i \in \mathcal{N}$
            \State Update $\boldsymbol{A}(t+1)$; calculate $\Phi_2(t+1)$

            \renewcommand{\algorithmicthen}{} 
            \If{} 
                \State $|\Phi_2(t+1)-\Phi_2(t)| < \varepsilon, \quad \textit{converged} \leftarrow \text{true}$
            \EndIf
            \renewcommand{\algorithmicthen}{\textbf{then}} 
            
            \State $t \leftarrow t+1$
        \EndWhile
        
        \State \Return $\mathbf{Q}^*$, $\mathbf{P}^*$, $\mathbf{C}^*$
    \end{algorithmic}
\end{algorithm}

When only the strategy of UAV $i$ changes from $\Gamma_{i2}$ to $\Gamma_{i2}^{\prime}$, the utilities of other UAVs remain unchanged. The change in the potential function is derived solely from the change in $\rho_{i2}$. Since $\Gamma_{-i2}$ is unchanged, only the terms of $\rho_{i2}$ change:
\begin{equation}
\Delta\Phi_2{=}\rho_{i2}\left(\Gamma_{i2}^{\prime},\Gamma_{-i2}\right){-}\rho_{i2}(\Gamma_{i2},\Gamma_{-i2}){=}\Delta\rho_{i2}.
\end{equation}

The proofs that $\Delta\Phi_1{=}\Delta\rho_{i1}$ and $\Delta\Phi_2{=}\Delta\rho_{i2}$ demonstrate that the change in the utility function caused by a unilateral action change of each UAV is identical to the change in the potential function. Additionally, according to the definition of EPGs, both $\mathscr{F}_{1}$ and $\mathscr{F}_{2}$ constructed in this paper are EPGs. This guaranties that $\mathscr{F}_{1}$ and $\mathscr{F}_{2}$ each possess at least one NE state, ensuring that $\mathscr{F}_{1}$ and $\mathscr{F}_{2}$ can each converge to a stable and feasible strategy state during the iteration process, and the final optimization result will not fluctuate continuously due to unilateral changes by individual UAVs.

\section{LLM-based EPG for Function and Weight Generating}
\label{sec:6}
In Section\ref{sec:5}, we established a distributed optimization mechanism based on EPGs. However, the utility functions and their corresponding weight coefficients are highly sensitive to dynamic network environments, making manual parameter tuning empirically difficult. To address this, this section introduces a RAG-assisted LLM framework to automate the utility design. 

As illustrated in Fig. \ref{fig:rag_framework}, the proposed RAG-based LLM optimization framework follows a closed-loop serialized logic consisting of four modules. First, in the knowledge vector database module, multi-source domain knowledge such as academic papers and protocols is segmented and embedded into a vector space. Second, the knowledge retrieval module receives the network manager intent and performs a cosine similarity search to extract relevant technical fragments. Third, the result generation module feeds these contexts into the LLM as structured prompts to formulate initial utility functions. Fourth, the iterative refinement module allows the manager to provide feedback, enabling the LLM to fine-tune the utility weights such as $\eta$ and $\psi$ until the generated parameters satisfy the specific deployment requirements of the current scenario. This work improves the algorithm adaptability and generalization ability under different network configurations.
\subsection{Knowledge Base Construction}

In this paper, the knowledge system of the pre-constructed knowledge base is extended by integrating an external knowledge base, such as academic papers related to UAV resource optimization from IEEE Xplore. The final knowledge base consists of the pre-constructed knowledge base and external academic papers. To achieve efficient semantic retrieval, these knowledge bases in raw text format must be converted into machine-understandable vector representations. 

The construction of a knowledge base requires transforming inputs into semantic information to facilitate subsequent retrieval and generation processes. Specifically, for pre-constructed knowledge base documents or academic texts denoted as $\mathrm{U}$ from various sources, an equidistant segmentation method is employed to partition the text into a set of knowledge blocks $\mathrm{U}=\{u_1, \dots, u_o, \dots, u_O\}$. Consequently, the vector representation of a knowledge block $u_o \in \mathrm{U}$ is given by:

\begin{equation}
    \mathcal{E}(u_o) = \text{Embed}(u_o).
\end{equation}

where $\mathcal{E}(\cdot)$ denotes the process of representing a knowledge block as a vector utilizing the text embedding model $\text{Embed}(\cdot)$. Possessing robust representation capabilities, the text embedding model can encode natural language into vector formats processable by neural networks, thereby playing a pivotal role in realizing human-machine interaction. In this paper, the text-embedding-ada-002 model is utilized as the text embedding model \cite{openai2022embedding}. Subsequently, the vector representation $\mathcal{E}(u_o)$ of the block is stored in a vector database. The knowledge vectors of all texts collectively constitute the knowledge vector database, facilitating the subsequent RAG process.

This paper employs RAG for document retrieval and generation. The RAG process involves retrieving relevant domain knowledge from the constructed knowledge vector database to support the LLM in generating the mathematical models required by the UAVN deployment constructor. The RAG process acquires knowledge by inputting the construction requirement $q$, and subsequently utilizes the retrieved knowledge as a prompt to assist the LLM in generating the modeling solution.

\subsection{Retrieval Knowledge Base}

The retrieval model $\mathrm{U}_{\mathfrak{o}}(u_{\mathrm{TOP}-\mathcal{K}}|\varrho)$ is based on a dual-encoder architecture for semantic retrieval, meaning the semantic encoding of the construction requirement $\varrho$ and the semantic encoding of the knowledge chunks $U$ are used to retrieve the most relevant knowledge chunks according to the construction requirement $\varrho$ to support mathematical modeling. The retrieval model is expressed as:
\begin{equation}
\mathrm{U}_{\mathfrak{o}}(u_{\mathrm{TOP}-\mathcal{K}}|\varrho)=\underset{\mathcal{K}}{\operatorname*{\operatorname*{argmax}}}\left(CS\left(\mathcal{E}(u_o),\mathcal{E}(\varrho)\right)\right),o\in O,
\end{equation}
where $\mathrm{U}_{\mathfrak{o}}(u_{\mathrm{TOP}-\mathcal{K}}|\varrho)$ denotes the truncation of the original probability distribution using the top-$\mathcal{K}$ technique, i.e., selecting the top $\mathcal{K}$ most relevant knowledge chunks based on the construction requirement. $\mathcal{E}_\varrho$ is the vector representation of the construction requirement, and $CS(\mathcal{E}_{u_o}, \mathcal{E}_\varrho)$ represents the cosine similarity between the construction requirement and the knowledge chunk $\mathrm{u}_{\mathfrak{o}}$.

Cosine similarity is utilized to measure the semantic similarity between two vectors. The semantic similarity between vector $\mathcal{E}_{u_o}$ and vector $\mathcal{E}_\varrho$ is expressed as:
\begin{equation}
\mathrm{CS}\left(\mathcal{E}(u_o),\mathcal{E}(\varrho)\right)=\frac{\sum_{o=l}^O\left(E(\varrho)[\ell]{\times}E(u_o)[\ell]\right)}{\sqrt{\sum_{\ell=l}^\mathcal{L}\left(E(\varrho)[\ell]\right)^2}\sqrt{\sum_{o=l}^O\left(E(u_o)[\ell]\right)^2}},
\end{equation}
where ${\ell}$ represents the vector index and $\mathcal{L}$ represents the length of the vector. From this, the semantic similarity between the construction requirement $\varrho$ and the knowledge chunk $u_o$ can be obtained.

\subsection{Result Generation}

In this paper, retrieved knowledge chunks are directly used as prompts input to the generation model. The function of the generation module is to generate a correct response based on the given construction requirement $\varrho$ and the corresponding retrieved knowledge $u_{\text{TOP}-\mathcal{K}}$. The generation model $\mathrm{U}_\Theta(\Xi\mid\varrho,u_{\mathrm{TOP}-\mathcal{K}})$ is expressed as follows:
\begin{equation}
\mathrm{U}_{\Theta}(\mathbb{S} \mid \mathcal{\varrho}, u_{\text{TOP}-\mathcal{K}}) = \prod_{\mathbb{I}}^{\mathbb{L}} p_{\theta}(\mathbb{S}_{\mathbb{I}} \mid \mathcal{\varrho}, u_{\text{TOP}-\mathcal{K}}, \mathbb{S}_{\mathit{1}:\mathbb{I}-\mathit{1}}),
\end{equation}
where $\mathbb{S}$ represents the generated response sequence of length $\mathbb{L}$, and the output of each token $\mathbb{S}_\mathbb{I}$ depends on the preceding $\mathbb{I}-1$ tokens $\mathbb{S}_{\mathit{1}:\mathbb{I}-\mathit{1}}$, the construction requirement $\varrho$, and the retrieved knowledge chunk $u_{\text{TOP}-\mathcal{K}}$.

The generation module is serviced by any LLM, such as ChatGPT\cite{singh2023chat}, Grok\cite{xai2024grok}, or DeepSeek\cite{deepseek2024llm}. In this paper, a pre-trained GPT is employed to implement the model $\mathrm{U}_{\Theta}(\mathbb{S} \mid \mathcal{\varrho}, u_{\text{TOP}-\mathcal{K}})$, utilizing retrieved knowledge and construction requirements as input prompts to generate the utility functions and their weight coefficients for the EPG.

The setting of weight coefficients for the utility functions in the L3-EPG and AG-EPG algorithms suffers from reliance on manual experience and a lack of environmental adaptability. To address these challenges, a utility function weight coefficient generation module based on LLMs and a self-built knowledge base is designed to achieve the autonomous generation of utility functions and their weight coefficients. This method is executed only during the initialization phase and does not alter the EPG framework. Specifically, the system first collects parameters such as the number of UAV nodes and their coordinates, GU distribution, communication radius, interference power, and energy budget, which are converted into structured query prompts and input into the LLM. Subsequently, based on the self-built knowledge base containing system models, simulation results, and historical weight records, the model outputs weight coefficients adapted to the current scenario. Moreover, logical verification is performed to confirm whether the weights satisfy connectivity, utility-potential function consistency, and energy constraints. If the verification passes, the weights are written into the algorithm; otherwise, the system reverts to historical valid weights and records the deviation information in the knowledge base. Upon completion of optimization, weight adjustment information and scenario parameters are appended to the knowledge base as key-value pairs. When the knowledge base reaches a threshold, parameter-efficient fine-tuning is performed on the LLM. This operation updates only the weight generation and recalls similar historical records via vector retrieval, improving subsequent generation accuracy.

As shown in Fig. \ref{fig:rag_framework}, the generation of weight coefficients employs a human-machine collaborative interaction method. First, the network builder inputs scenario parameters, including communication area range, number of UAV nodes and coordinates, GU equipment distribution, communication radius, average communication interference power level, etc., and defines the optimization objectives. Subsequently, the LLM generates initial utility functions and their weight coefficients based on the self-built knowledge base. Next, the LLM confirms with the network builder whether the generated utility functions meet actual requirements. If the network builder deems that the weight coefficients need adjustment, the LLM regenerates corrected weight coefficients by combining knowledge base information with feedback opinions. This process iterates until the network builder stops the interaction. After the interaction is completed, the determined weight coefficients and corresponding scenario parameters are stored in the knowledge base as new samples for reference in subsequent scenarios. Once the utility functions and their weight coefficients are generated, the UAVs execute the L3-EPG and AG-EPG two-stage optimization algorithms, finally outputting UAVN deployment optimization decisions that meet performance requirements.

\begin{figure}[htbp]
    \centering
    \includegraphics[width=\columnwidth]{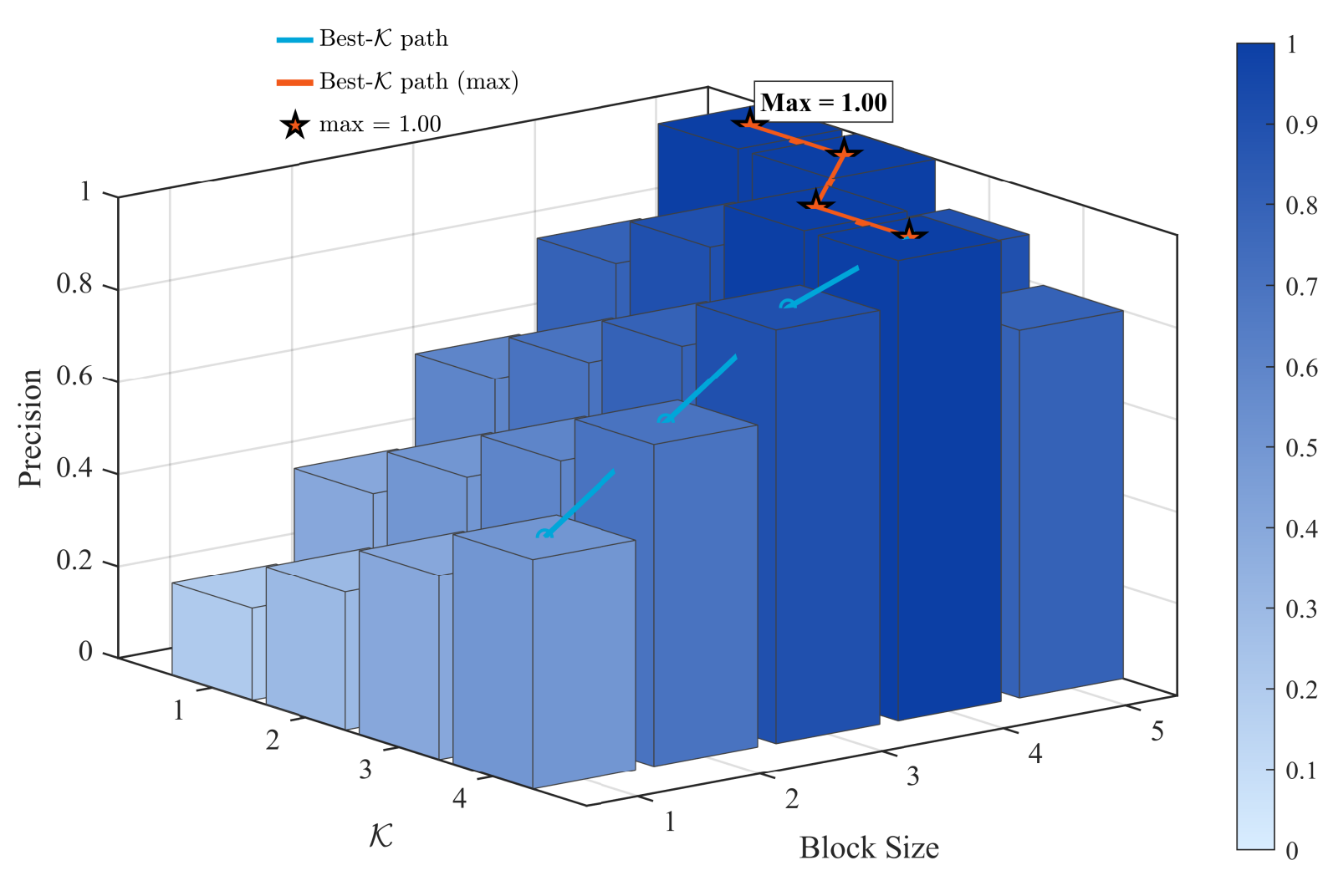} 
    \caption{Retrieval precision under different knowledge block sizes.}
    \label{fig:block_size_retrieval}
\end{figure}

\begin{figure}[htbp]
    \centering
    \subfloat[Consistency of L3-EPG]{
        \includegraphics[width=0.46\columnwidth]{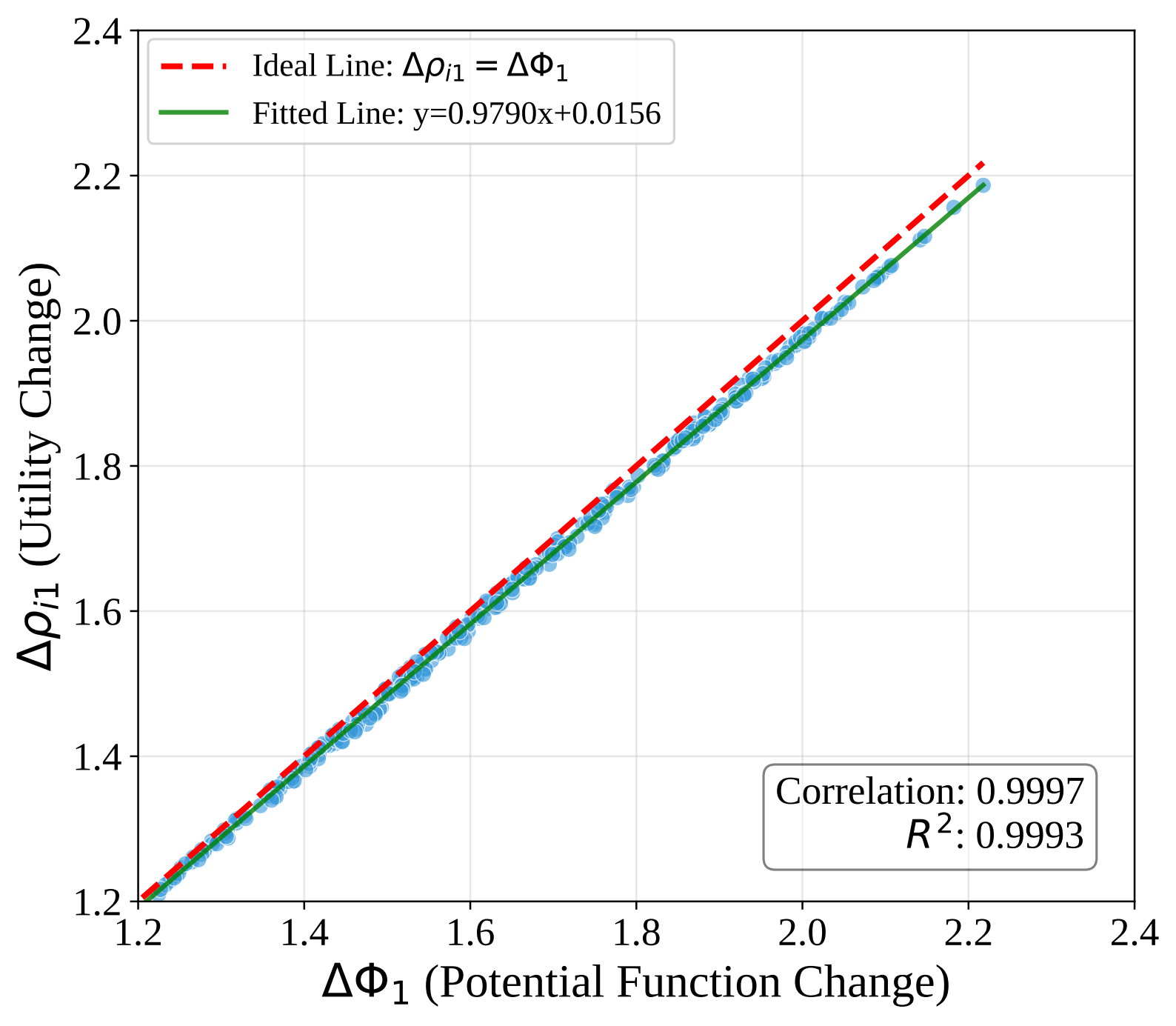}
        \label{fig:convergence_a}
    }
    \hfil 
    \subfloat[Consistency of AG-EPG]{
        \includegraphics[width=0.46\columnwidth]{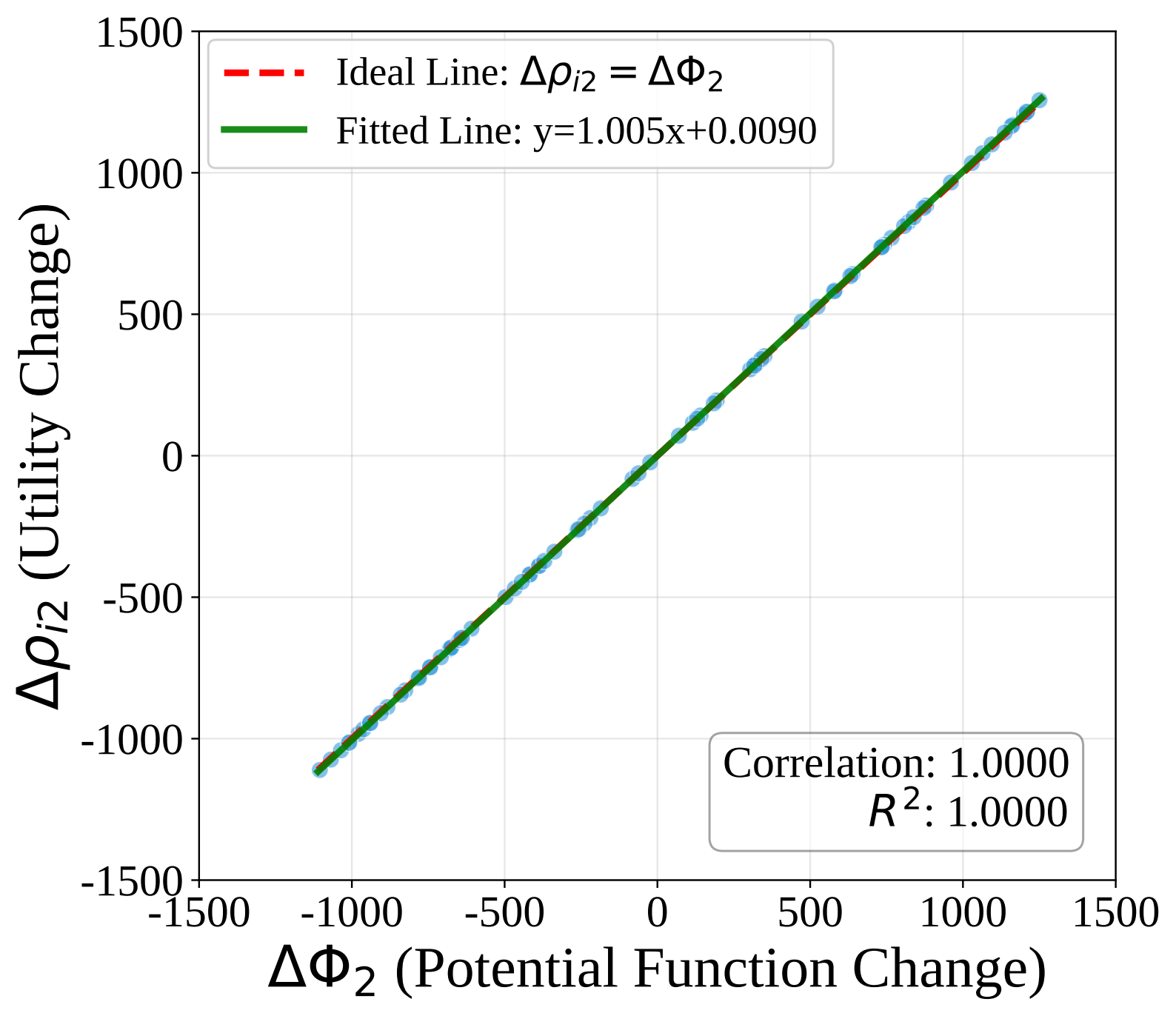}
        \label{fig:convergence_b}
    }
    \\ 
    \vspace{0.2cm} 
    \subfloat[Convergence of L3-EPG]{
        \includegraphics[width=0.46\columnwidth]{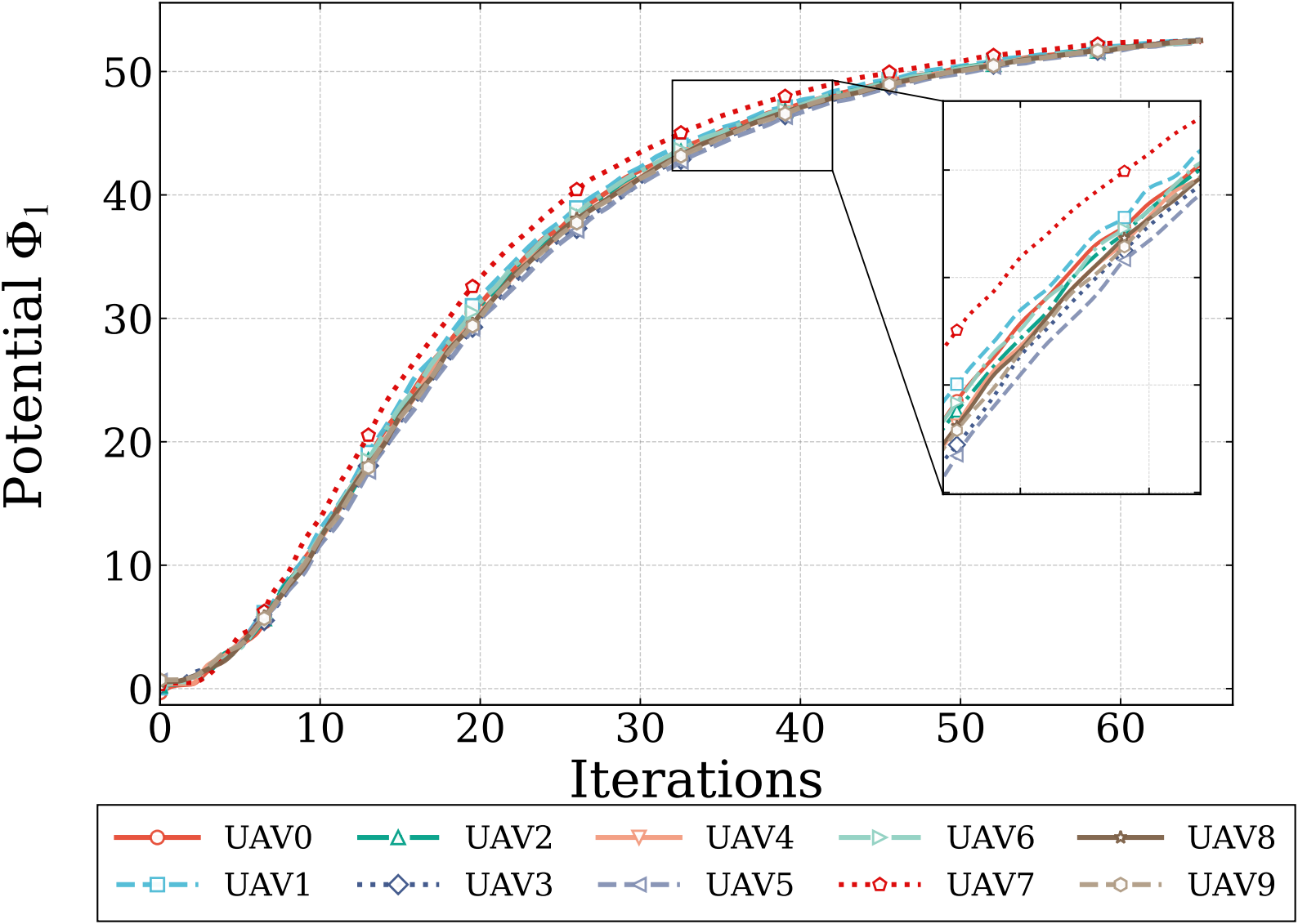}
        \label{fig:convergence_c}
    }
    \hfil
    \subfloat[Convergence of AG-EPG]{
        \includegraphics[width=0.46\columnwidth]{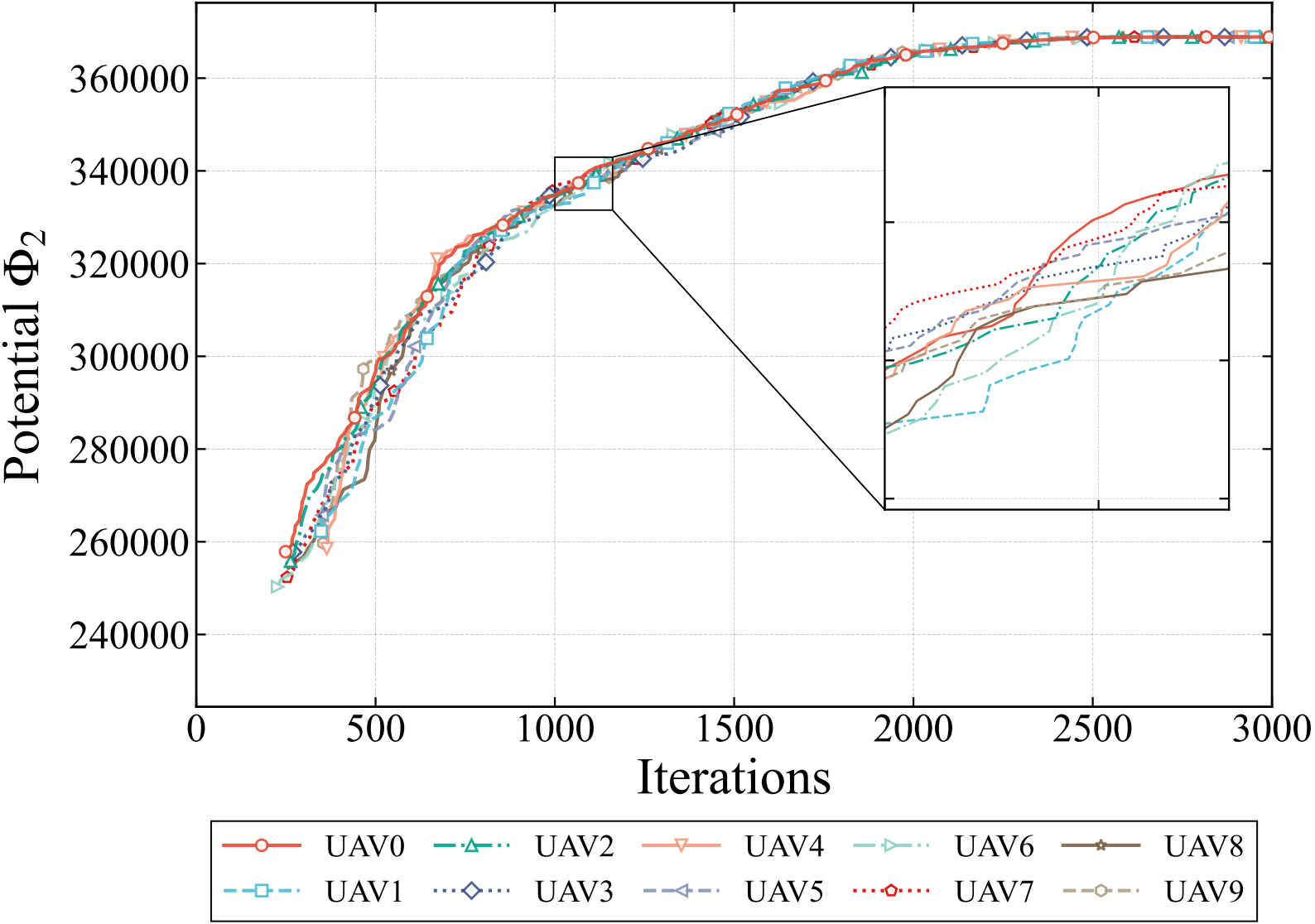}
        \label{fig:convergence_d}
    }
    \caption{Verification of consistency and convergence of algorithms.}
    \label{fig:convergence_main} 
\end{figure}

\section{Experimental Results and Analysis}
\label{sec:7}
To comprehensively validate the effectiveness of the proposed framework, the remainder of this section is organized as follows: Section\ref{sec:7}-A evaluates the retrieval precision of the RAG module. Section\ref{sec:7}-B verifies the convergence and global consistency of the EPG algorithms. Section\ref{sec:7}-C visualizes the dynamic deployment optimization process, and Section\ref{sec:7}-D compares the overall network performance against the aforementioned benchmark algorithms.

The simulation experiment is conducted in a Python 3.8 environment on a high-performance server equipped with an NVIDIA RTX 4090 GPU. The experiment constructs a low-altitude communication network scenario comprising 10 UAVs and 20 GUs. To ensure an unbiased evaluation of network coverage and throughput, the initial coordinates of the UAV nodes and GUs are generated within an 10 km $\times$ 10 km area according to a uniform distribution, utilizing a fixed random seed to guarantee experimental reproducibility. For the LLM-enhanced module, the GPT-4-turbo model is utilized as the reasoning engine, integrated with the text-embedding-ada-002 model for semantic retrieval. Based on the optimization results in Section\ref{sec:7}-A, the RAG block size is set to 4 and the retrieval count is set to 3 to achieve peak retrieval precision. In the AG-EPG phase, the convergence threshold is set to $10^{-4}$ to ensure high precision. The LoS probability, denoted as $P_{\text{LoS}}(\theta)$, for the U2G links adopts the elevation-angle-dependent model from ITU-R P.1410 \cite{itu2019propagation}. The environmental parameters are set to $\alpha=9.6$ and $\beta=0.28$. The system bandwidth is $B=2$ MHz, the path loss exponent is $n_0=2.5$, the noise power spectral density is $N_0=-174$ dBm/Hz, and the antenna gains are $G_t, G_r=3$ dBi. Other simulation parameters are summarized in Table \ref{table:simulation_parameters}.

\begin{table}[htbp]
\centering
\caption{Simulation parameters}
\label{table:simulation_parameters}
\begin{tabular}{l|l|l}
\hline\hline
\textbf{Symbol} & \textbf{Meaning} & \textbf{Value} \\ \hline
$N$ & Number of UAVs & 10 \\ \hline
$M$ & Number of GUs & 20 \\ \hline
$\mathbb{R}^3$ & Communication area range & $10 \text{ km} \times 10 \text{ km}$ \\ \hline
$[z_{\min}, z_{\max}]$ & UAV altitude range & $[100 \text{ m}, 300 \text{ m}]$ \\ \hline
$R_c$ & Maximum communication radius & $4000 \text{ m}$ \\ \hline
$B$ & System bandwidth & $2 \text{ MHz}$ \\ \hline
$p_{\max}$ & Maximum transmit power & $2.0 \text{ W}$ \\ \hline
$p_{\min}$ & Minimum transmit power & $0.5 \text{ W}$ \\ \hline
$n_0$ & Path loss exponent & 2.5 \\ \hline
$N_0$ & Noise power spectral density & $-174 \text{ dBm/Hz}$ \\ \hline
$G_t, G_r$ & Antenna gain & $3 \text{ dBi}$ \\ \hline\hline
\end{tabular}
\end{table}

To validate the effectiveness of the proposed method, the algorithm is compared with four benchmark algorithms selected based on their technical representativeness and comparability to the proposed framework.

\begin{itemize}
    \item \textbf{EPG based on Best Response Dynamics (BRD-EPG):} Operating within the EPG framework, this algorithm employs BRD. It iteratively updates UAV coordinates, transmit power, and user association via gradient descent, basing decisions on the maximization of each UAV's local utility \cite{tembine2010evolutionary}.
    
    \item \textbf{Non-Cooperative Game Solver based on BRD (BRD-NCG):} This method makes decisions within a non-cooperative game framework. It utilizes BRD to iteratively update strategies to reach a NE, where strategy updates depend solely on the local utility function of each individual UAV.
    
    \item \textbf{Evolutionary Game (ETG):} This approach employs Replicator Dynamics \cite{yang2023multi} to model the evolutionary process of strategies within a population. It adjusts the strategy probability distribution in an online manner. Its decision mechanism considers utility intensity and population evolutionary characteristics, dynamically updating strategies based on collective behavior.
    
    \item \textbf{Genetic Algorithm (GA):} GA optimizes continuous parameters through offline population evolution, where the fitness function is defined by the global objective function. Furthermore, GA utilizes crossover and mutation mechanisms to obtain parameter convergence estimates without considering distributed game constraints \cite{gao2024joint}.
\end{itemize}

\subsection{Analysis of LLM Knowledge Retrieval}

As shown in Fig. \ref{fig:block_size_retrieval}, the knowledge block size and the value of the retrieval count $\mathcal{K}$ significantly influence retrieval precision. When the block size is small and $\mathcal{K}$ is low, the overall precision is relatively poor. This is primarily because the volume of effective information returned during the retrieval stage is insufficient, leading to a lack of adequate grounding for the generation stage and, consequently, degrading retrieval precision. As the block size increases and $\mathcal{K}$ is moderately raised, the retrieved results can provide more complete context support, thereby improving precision. However, when the block size continues to increase or $\mathcal{K}$ becomes excessively high, a decline in precision is observed. The main reason is that the proportion of irrelevant or weakly relevant information in the retrieved content rises. With longer input texts, the LLM is required to filter key conditions and core clues from a larger scope, which increases susceptibility to issues such as unfocused key point extraction and insufficient utilization of key constraints, ultimately reducing precision. Therefore, it is evident that larger block sizes and $\mathcal{K}$ values are not necessarily better; instead, a balance must be struck between information coverage and redundancy control. Considering the requirements for both retrieval precision and the control of retrieval text scale, this paper adopts parameter settings that achieve peak precision while avoiding excessive redundancy, specifically setting the block size to 4 and $\mathcal{K}$ to 3.

\subsection{Convergence of EPG}
\label{sec:topology_evolution}
\begin{figure*}[htbp]
    \centering
    
    \subfloat[Time slot 0 (fully connected)]{
        \includegraphics[width=0.31\textwidth]{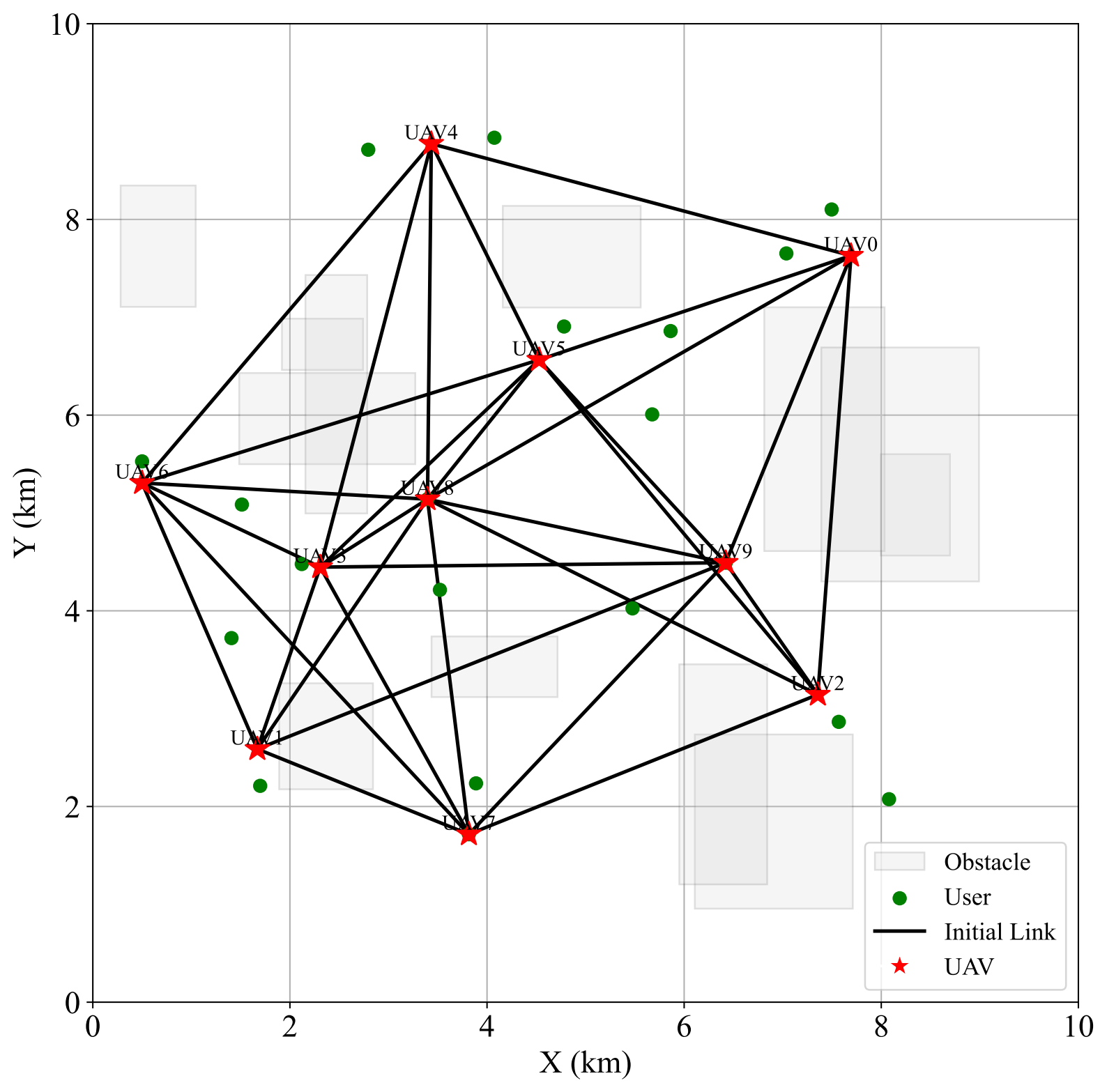}
        \label{fig:4-2a}
    }
    \hfil 
    \subfloat[Time slot $t$]{
        \includegraphics[width=0.31\textwidth]{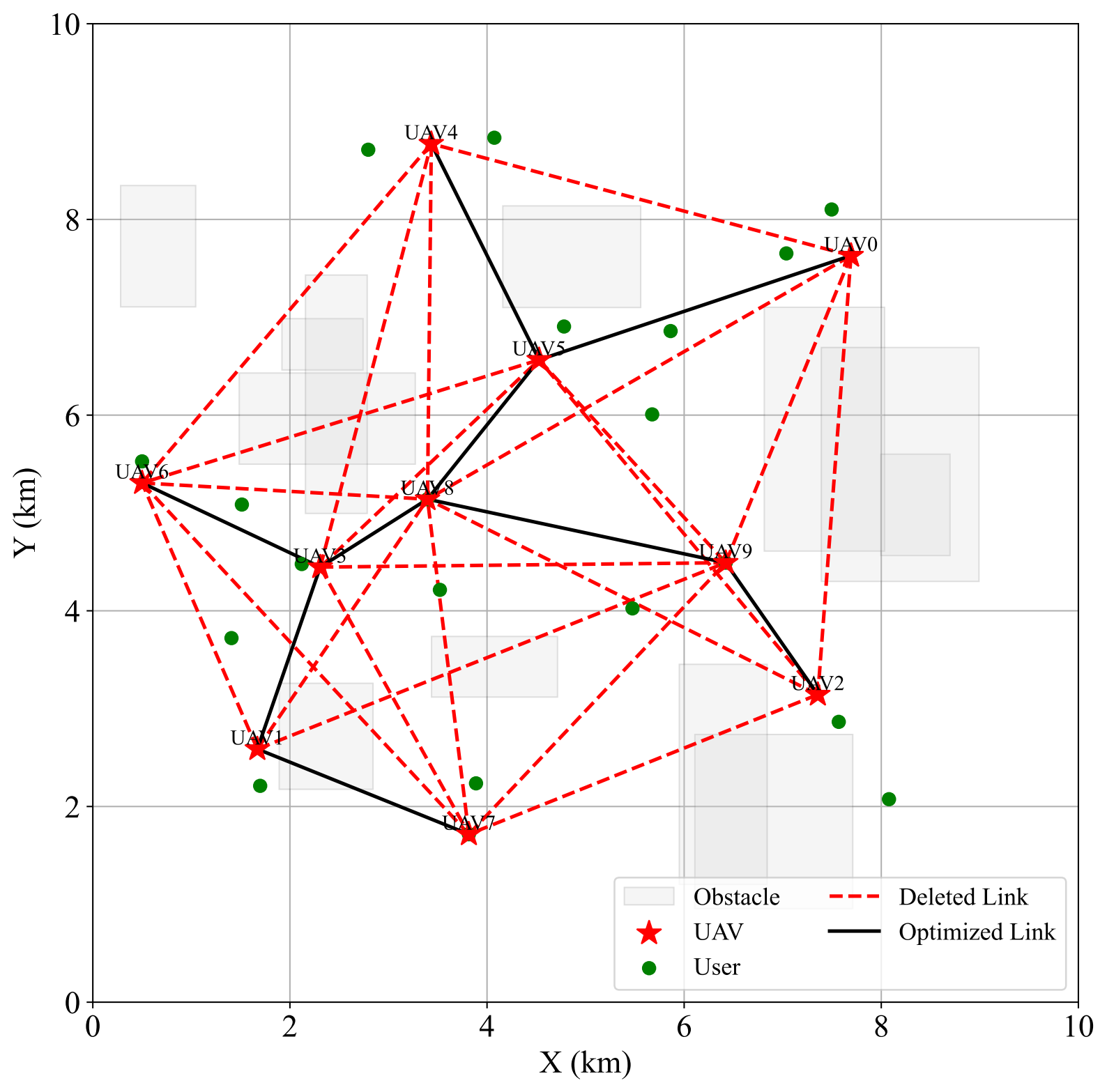}
        \label{fig:4-2b}
    }
    \hfil 
    \subfloat[Time slot $t+32$]{
        \includegraphics[width=0.31\textwidth]{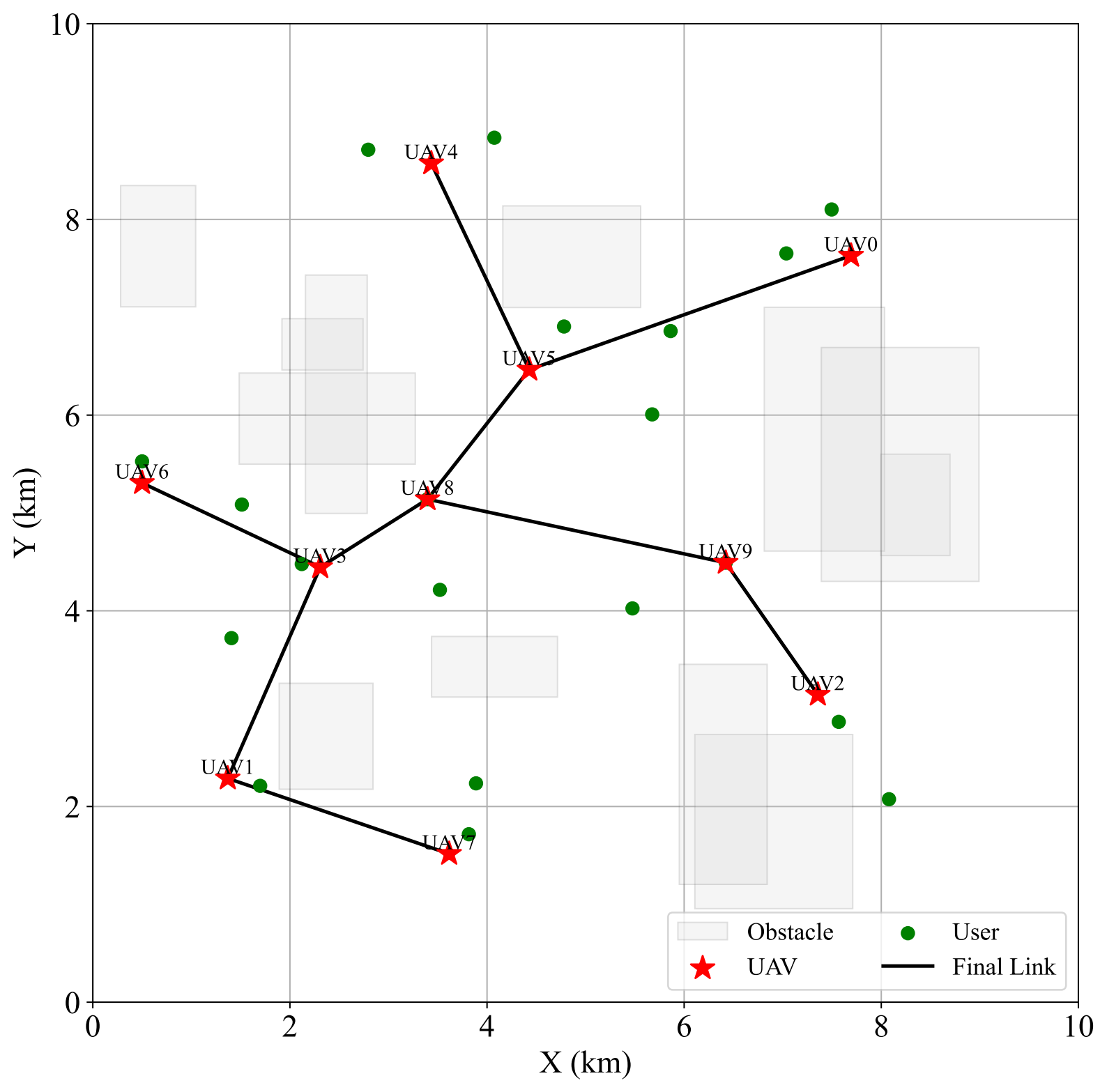}
        \label{fig:4-2c}
    }
    
    \caption{Topology optimization process.}
    \label{fig:4-2}
\end{figure*}

\begin{figure*}[htbp]
    \centering

    \subfloat[3D topology]{
        \includegraphics[width=0.48\textwidth]{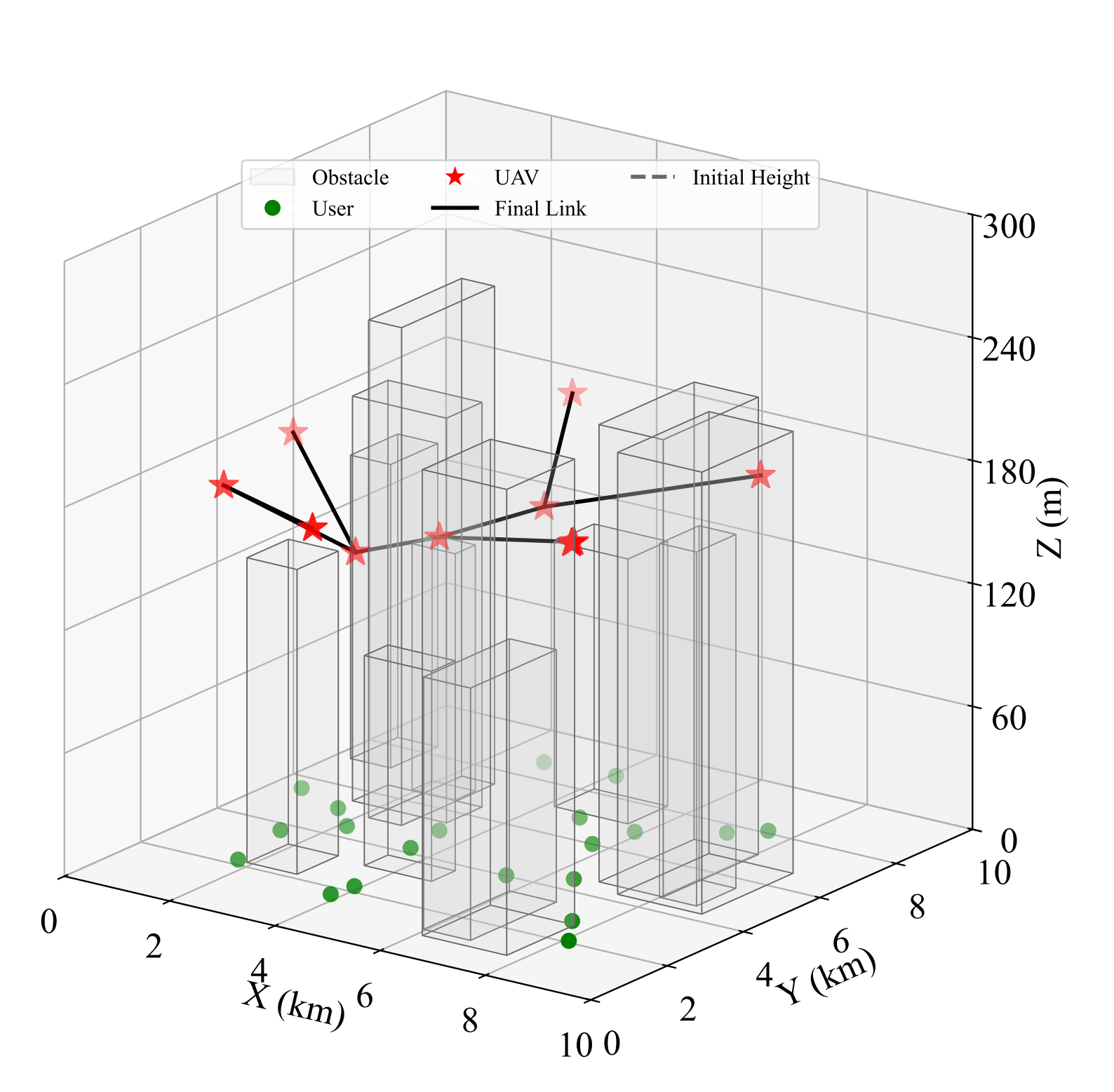} 
        \label{fig:4-3a}
    }
    \hfil %
    \subfloat[Node altitude]{
        \includegraphics[width=0.48\textwidth]{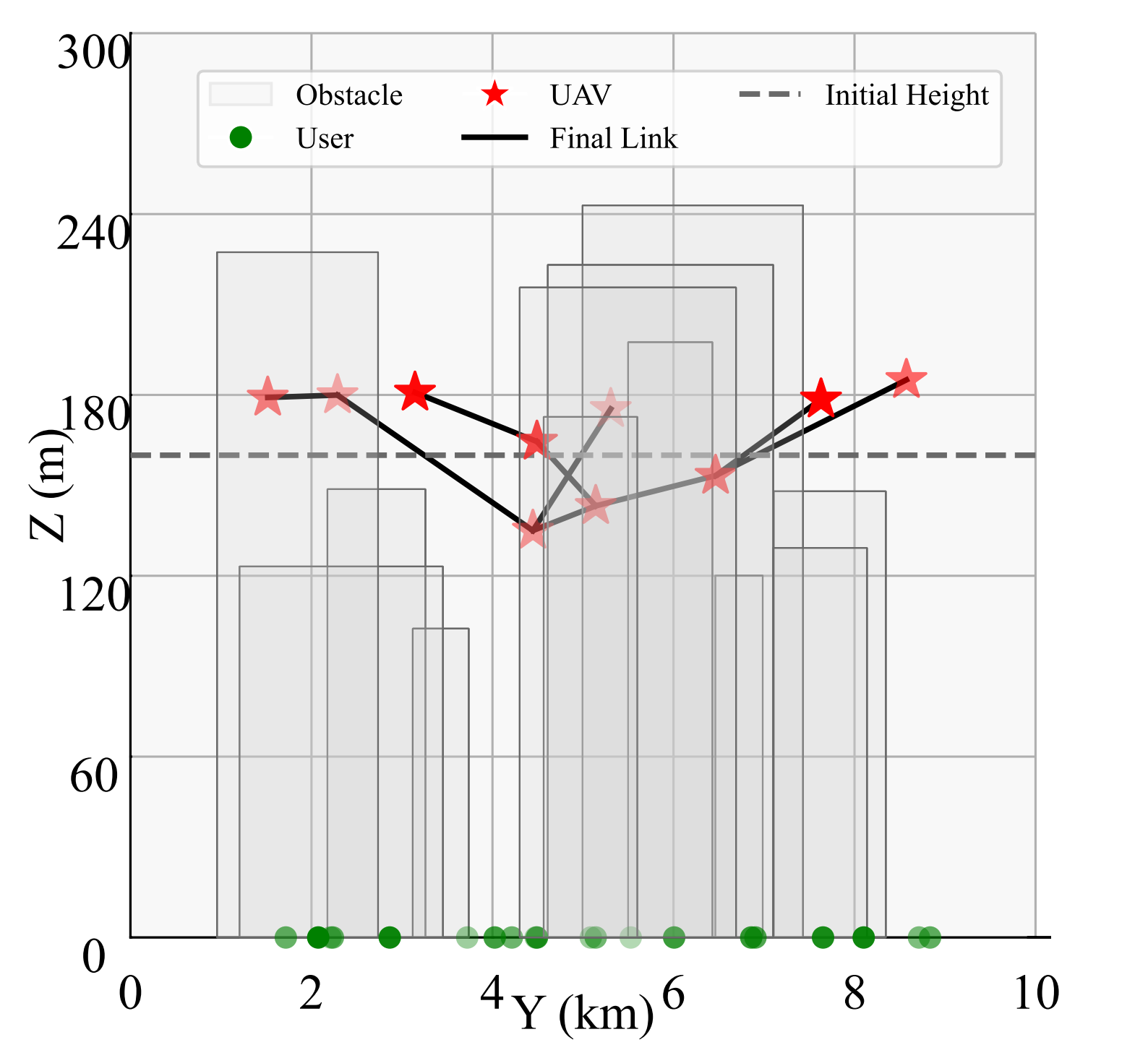}
        \label{fig:4-3b}
    }
    
    \caption{Spatial distribution of UAV nodes.}
    \label{fig:4-3}
\end{figure*}
Fig. \ref{fig:convergence_a} and Fig. \ref{fig:convergence_b} demonstrate the correspondence between changes in local utility and changes in the potential function, quantified using the correlation coefficient and the coefficient of determination ($R^2$). Experimental results indicate that the trends in the correlation coefficient and $R^2$ for both L3-EPG and AG-EPG are consistent and close. Fig. \ref{fig:convergence_c} and Fig. \ref{fig:convergence_d} further present the convergence trends of 10 UAVs during the iterative process. It can be observed that although the curves for L3-EPG and AG-EPG exhibit certain fluctuations in the early stages of iteration, their directions of change are highly consistent, characterized by a gradual rise from the initial stage to a final steady state. This convergence pattern, where individual trends are consistent and ultimately stable, indicates that the strategy updates of the algorithm not only enable individual UAVs to achieve improvements but also ensure that the update directions of different UAVs do not conflict with each other, demonstrating a feature of synergistic advancement in the overall iteration. Consequently, these results demonstrate that the changes in local utility brought about by unilateral updates remain highly consistent with the changes in the potential function, thereby avoiding the oscillatory situations common in distributed updates and verifying the stable convergence capability of the proposed method.

\subsection{Deployment Optimization}

Fig. \ref{fig:4-2} illustrates the topology evolution process during discrete link optimization. The initial topology exhibits dense links with numerous redundant connections. As shown in Fig. \ref{fig:4-2a} and Fig. \ref{fig:4-2b}, as the L3-EPG iteration proceeds, redundant links are progressively eliminated while satisfying connectivity constraints, resulting in a significant reduction in the number of links. This result aligns with the optimization objective of the discrete stage: to prioritize the retention of critical links while reducing unnecessary ones under the condition of guaranteeing network connectivity, thereby creating conditions for minimizing interference and energy consumption. Fig. \ref{fig:4-2c} displays the dynamically adjusted topology after a change in GU coordinates (at time slot $t+32$). It can be observed that the UAV coordinates have undergone adaptive adjustments, ensuring that all links successfully bypass or fly over obstacles.

Fig. \ref{fig:4-3a} visualizes the 3D structure of the network topology. Simultaneously, as observed in Fig. \ref{fig:4-3b}, which displays the flight altitudes of network nodes, the flight altitudes of certain UAVs are significantly higher than the initial altitude baseline. Their coordinates correspond to adjacent high-rise buildings, indicating that the continuous stage provides more adequate LoS conditions for critical connections by increasing local altitudes, thereby reducing the occurrence of obstructed links. Meanwhile, the altitudes of the remaining UAVs mostly remain near the initial altitude, with slight descents to reduce network communication distances, which is conducive to lowering communication latency. These results reflect the complementarity between the L3-EPG and AG-EPG.

\subsection{Network Performance}

\begin{figure*}[htbp]
    \centering
    
    \subfloat[UAV energy consumption]{
        \includegraphics[width=0.31\textwidth]{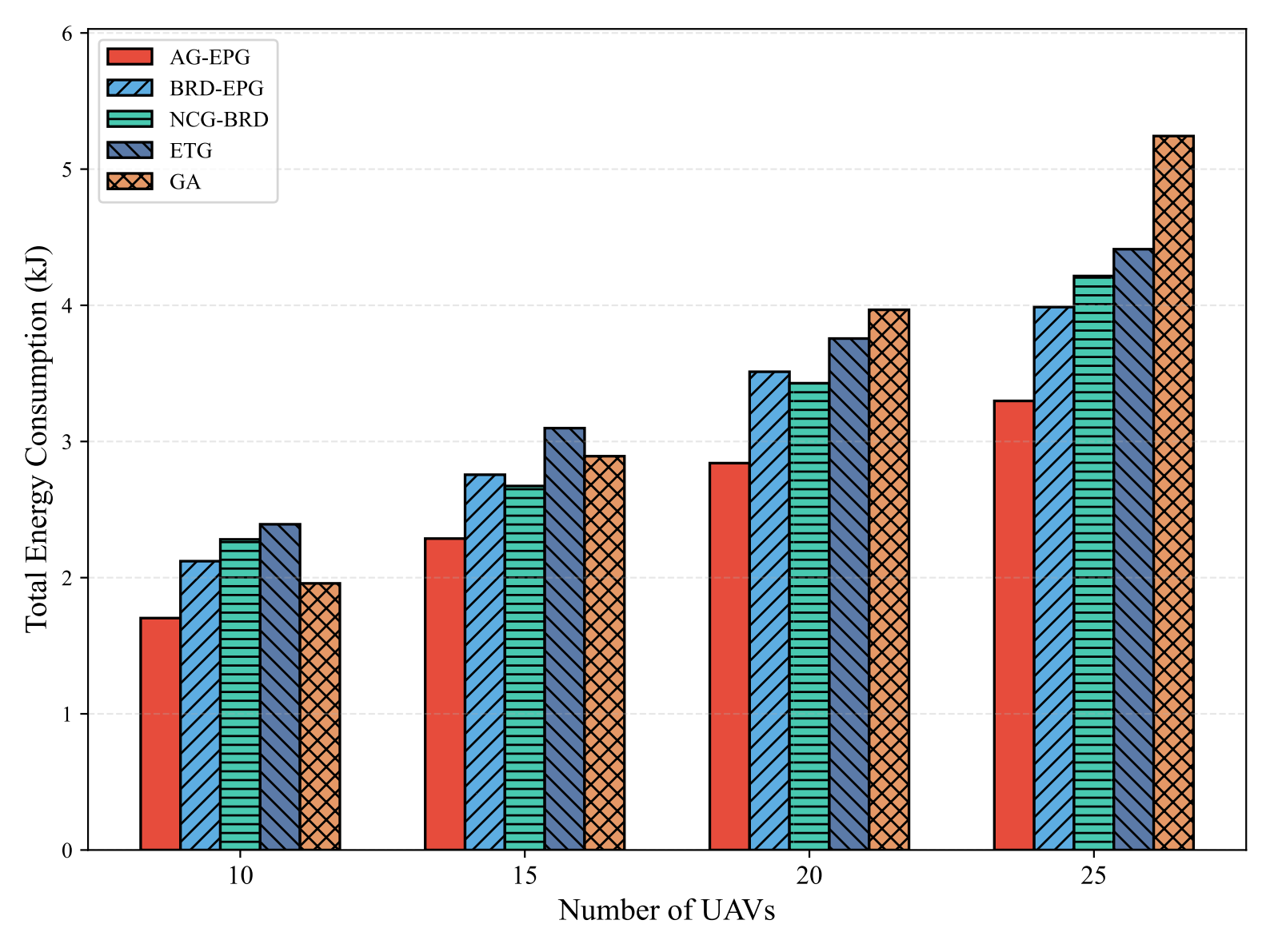}
        \label{fig:4-9a}
    }
    \hfil %
    \subfloat[Network latency]{
        \includegraphics[width=0.31\textwidth]{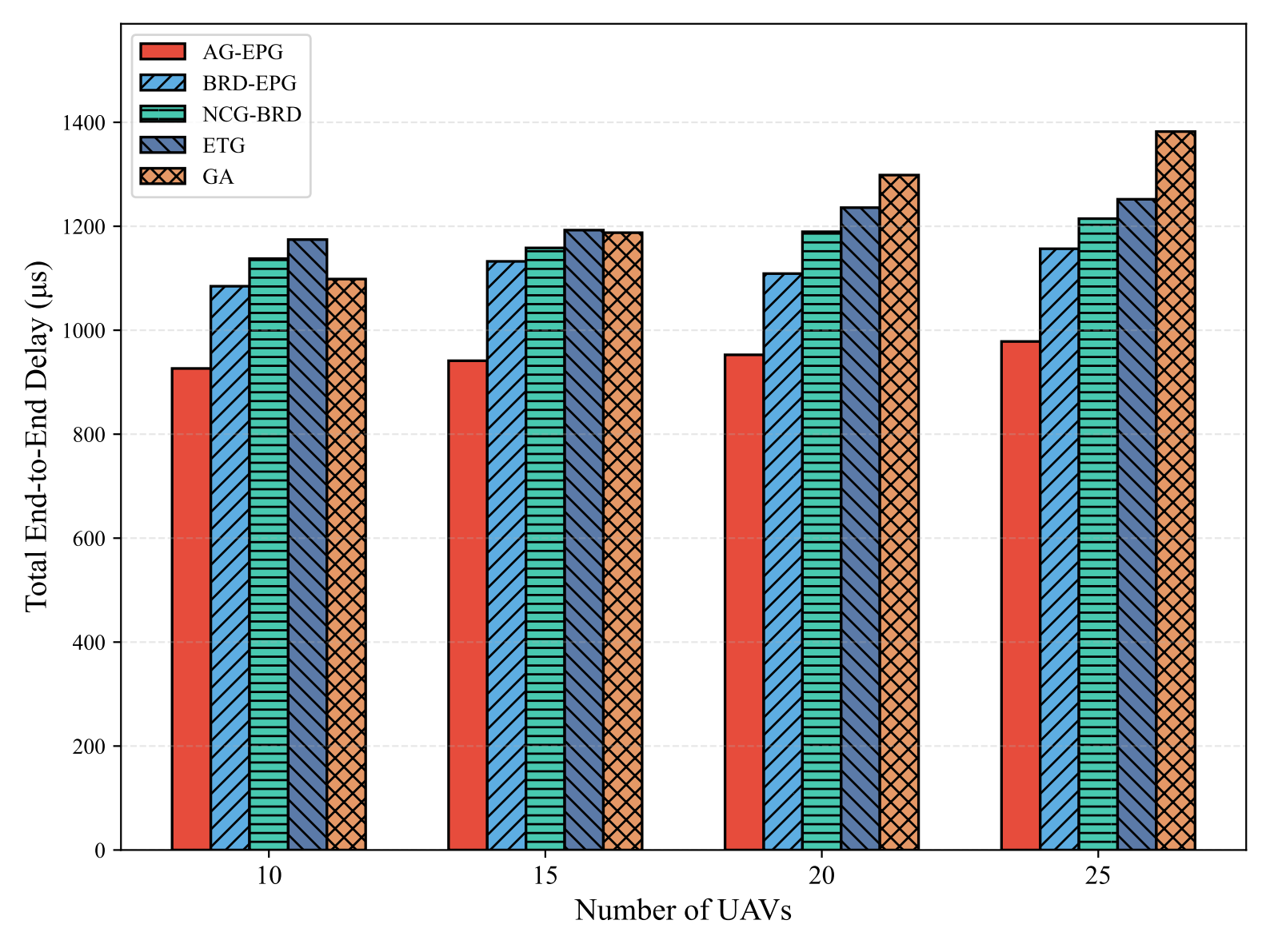}
        \label{fig:4-9b}
    }
    \hfil %
    \subfloat[Network throughput]{
        \includegraphics[width=0.31\textwidth]{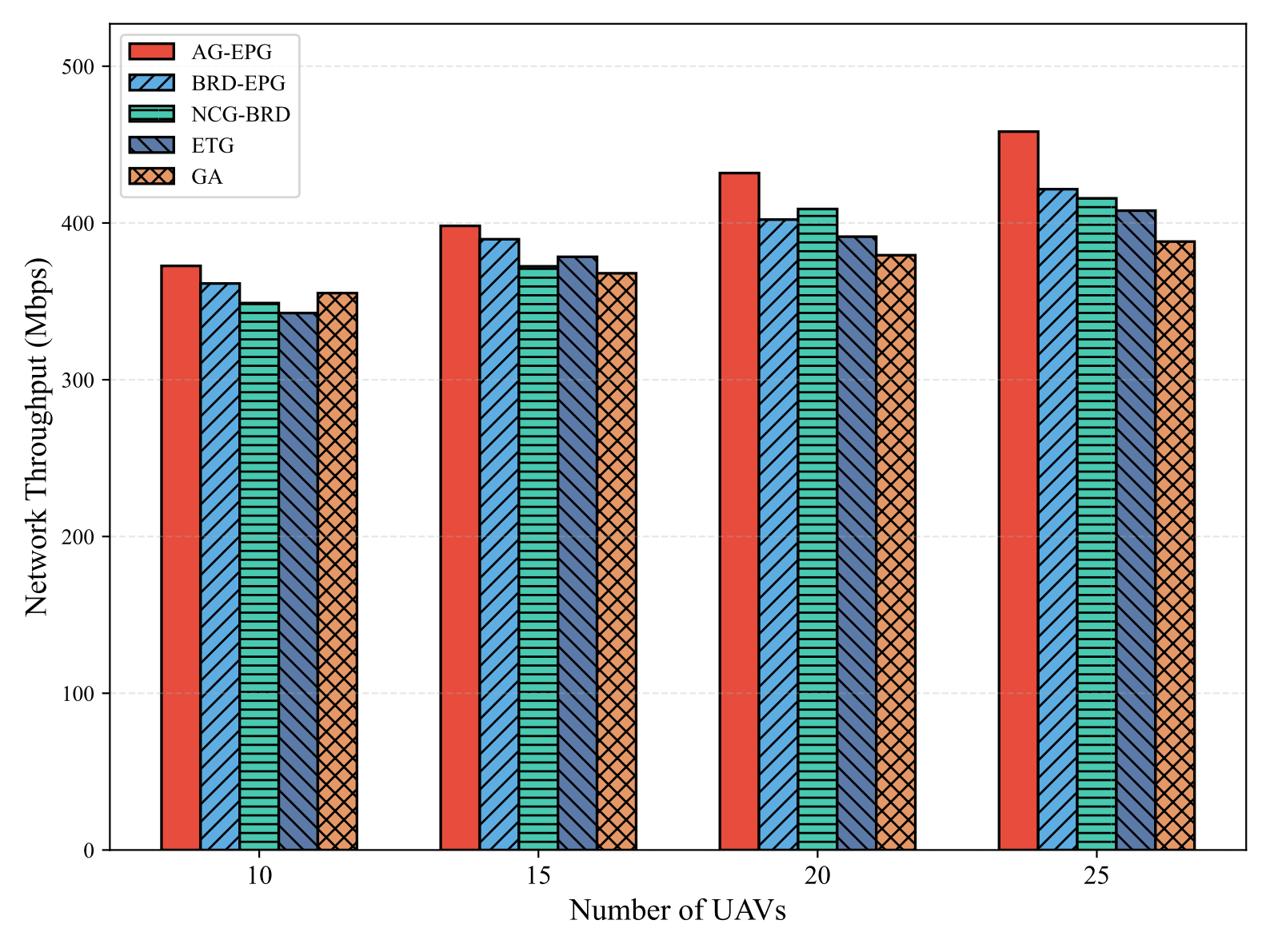}
        \label{fig:4-9c}
    }
    
    \caption{Comparison of network performance.}
    \label{fig:4-9}
\end{figure*}

Fig. \ref{fig:4-9} presents a comparison of energy consumption, latency, and network throughput among different algorithms under varying numbers of UAVs. AG-EPG achieves the highest throughput while maintaining lower total energy consumption and total latency across different UAV quantities. As the number of UAVs increases, the total energy consumption of all algorithms shows an upward trend, as shown in Fig. \ref{fig:4-9a}. Similarly, the total latency of all algorithms exhibits a slight upward trend with the increase in UAVs, as shown in Fig. \ref{fig:4-9b}. This is because maintaining network links and communication overhead increases correspondingly with the number of UAVs. However, since AG-EPG can reduce unnecessary connections during the topology optimization stage, the network energy consumption and latency based on AG-EPG remain the lowest. Furthermore, in the continuous optimization stage, by adjusting the 3D coordinates and transmit power of the UAVs, the communication distances of some links are reduced, and transmit power is optimized, thereby improving link transmission conditions. Compared with the four benchmark algorithms under different network scales, the network throughput is improved by approximately 8.4\%, as shown in Fig. \ref{fig:4-9c}.

\section{Conclusion}
\label{sec:8}

The paper studied the deployment problem in UAVN, which involves coupled discrete link decisions and continuous resource configurations with a Agentic AI-driven framework. To address the resulting mixed-integer nonconvex formulation, a dual spatial-scale optimization framework based on EPGs was proposed. At the large spatial scale, L3-EPG was adopted to optimize network link configurations, while at the small spatial scale, AG-EPG were used to jointly optimize UAV deployment, transmission power, and GU association. In addition, an LLM-based mechanism was incorporated to assist utility weight generation, reducing manual parameter tuning and improving adaptability across different scenarios. Despite the significant improvements demonstrated by the proposed framework, several directions for future research remain to be explored. On the one hand, the transition from simulation to real-world deployment requires addressing the on-board computational constraints of UAVs, particularly when executing LLM-based reasoning in real-time. On the other hand, the current model can be extended to consider more complex environmental factors, such as extreme weather conditions and high-speed mobility in dense 3D urban canyons, to further evaluate the robustness of the EPG-based learning.

\bibliographystyle{IEEEtran}
\input{Main.bbl}

\end{document}